\documentclass{ejm}
\usepackage{hyperref}
\usepackage{graphicx}% Include figure files
\usepackage{amsmath,amssymb}
\usepackage{color}

\begin{document}

\title[Simulating Surfactant Spreading]{Simulating Surfactant Spreading: \\ Impact of a Physically Motivated \\Equation of State}

\author[Dina Sinclair, Rachel Levy and Karen Daniels]{%
 Dina Sinclair$\,^1$,\ns
 Rachel Levy$\,^1$\ns
\and
Karen E. Daniels$\,^2$
}

\affiliation{%
  $^1\,$Mathematics Department, Harvey Mudd College, Claremont, CA, USA\\
    email\textup{\nocorr: \texttt{levy@g.hmc.edu}}\\
  $^2\,$Department of Physics, North Carolina State University, Raleigh, NC USA}

\date{14 February 2016}
\pubyear{2016}
\volume{000}
\pagerange{\pageref{firstpage}--\pageref{lastpage}}

\label{firstpage}
\maketitle

\begin{abstract}
For more than two decades, a single model for the spreading of a surfactant-driven thin liquid film has dominated the applied mathematics literature on the subject. Recently, through the use of fluorescently-tagged lipids, it has become possible to make direct, quantitative comparisons between experiments and models. These comparisons have revealed two important discrepancies between simulations and experiments: the spatial distribution of the surfactant layer, and the timescale over which spreading occurs.  In this paper, we present numerical simulations that demonstrate the impact of the particular choice of the equation of state (EoS) relating the surfactant concentration to the surface tension.  Previous choices of the model EoS have been an ad-hoc decreasing function.  Here, we instead propose an empirically-motivated equation of state; this provides a route to resolving some discrepancies and raises new issues to be pursued in future experiments. In addition, we test the influence of the choice of initial conditions and values for the non-dimensional groups.  We demonstrate that the choice of EoS improves the agreement in surfactant distribution morphology between simulations and experiments, and impacts the dynamics of the simulations. The relevant feature of the EoS, the gradient, has distinct regions for empirically motivated choices, which suggests that future work will need to consider more than one timescale. We observe that the non-dimensional number controlling the relative importance of gravitational vs. capillary forces has a larger impact on the dynamics than the other non-dimensional groups.  Finally, we observe that the experimental approach of using a ring to contain the surfactant could affect the surfactant and fluid dynamics if it disrupts the intended initial surfactant distribution. However, the meniscus itself does not significantly affect the dynamics. 
\end{abstract}


\maketitle

%================================================================
\section{Introduction \label{sec:intro}}

Chemicals that lower the surface tension of a fluid are known as surfactants (shorthand for surface active agents). The ability to predict and control the rate and extent to which surfactants spread over the surface of a fluid is important to improving their use in many applications. For example, they are present in healthy lungs to enable breathing, and are also used in industrial applications as stabilizers and dispersants \cite{karsa1999industrial}.
In human lungs, issues such as airway closure and reopening \cite{otis1993role} and the dynamics of mucus in the airway system \cite{craster2000surfactant,halpern1993surfactant,bull2003surfactant, levy2014pulmonary} are known to be tied to the presence of surfactants. 
In particular, biomedical engineers and applied mathematicians studying the liquid lining of the lungs of premature infants proposed a compelling model starting with a well-known thin film equation  and coupling the film to the surfactant through surface stress \cite{gaver1990,Gaver-1992-DST}.  The stress on the fluid is created by concentration gradients in the layer of insoluble surfactant (and thus a surface tension gradient in the fluid).  This, in turn, induces transport of that surfactant on the surface as the fluid moves \cite{jensen1992insoluble, jensen1993spreading}.

The model takes the form of two fourth-order nonlinear parabolic-hyperbolic partial differential equations. Mathematical interest in these model equations have led to several fruitful approaches to solutions such as asymptotics, similarity solutions, and numerical simulations \cite{levy2006motion,espinosa1993spreading, angelini2009bacillus,jensen1994self}.  The solutions provide predictions of spreading behavior including a spreading timescale as well as fluid and surfactant spatial distribution over time.  To test the model, experiments by physicists \cite{Fallest-2010-FVS, Strickland2014} have provided new measurements of the motion of both the surfactant molecules and the underlying fluid, directly testing the  validity of the model. While this has led to the identification of some limitations in the model \cite{swanson2014surfactant}, the new data also suggests possible improvements. 

This paper provides the first attempt to introduce evidence from the experiments back into the model.
For simplicity, we focus on two simple spreading geometries which we will call {\it outward-spreading}  \cite{Fallest-2010-FVS} and {\it inward-spreading} \cite{Strickland2014}. In both sets of experiments, the system starts with a uniform, millimetric film of glycerol placed on a silicon wafer within a cylindrical container. A smaller retaining ring is placed at the surface of the fluid, in the center of the container. In the outward-spreading experiments, the surfactant is placed inside the retaining ring so that it spreads outward once the ring is lifted. In the inward-spreading experiments, the surfactant is placed outside the ring and spreads inwards.  In simulations, these two cases will be implemented through analogous initial conditions.

\begin{figure}
\centering
\includegraphics[width=4in]{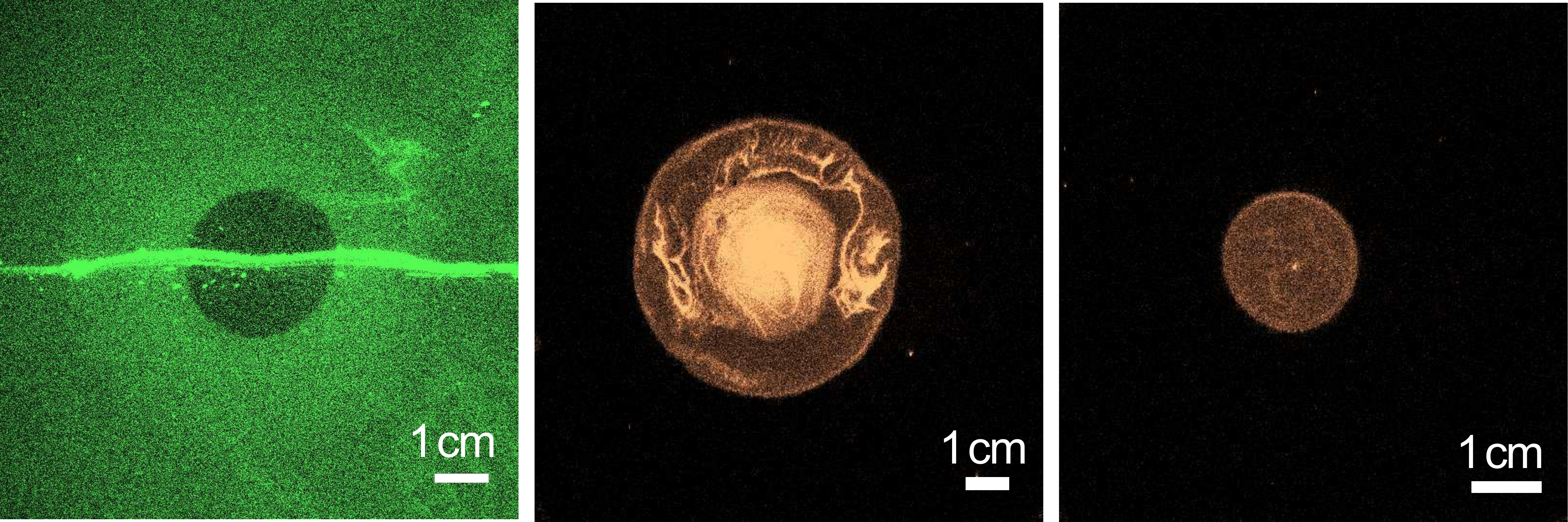}
\caption{Images of fluorescently-tagged surfactant spreading from experiments \cite{swanson2014surfactant,Strickland2014} viewed from above; brighter regions have a larger concentration of  surfactant. Left: inward-spreading below $\Gamma_c$ with laser line to show fluid profile with capillary ridges. Middle: outward spreading above $\Gamma_c$ with central ``reservoir'' region of high surfactant concentration. Right: outward spreading below $\Gamma_c$ with more uniform surfactant concentration. Figure adapted from \cite{swanson2014surfactant,Strickland2014}. }
\label{fig:exper}
\end{figure}

In both cases, it is not just the surfactant layer that moves:  the glycerol itself is advected by the surfactant, pulled toward regions with less surfactant (higher surface tension). In order to track the surface distribution of the surfactant molecules, experiments use a fluorescently tagged lipid called NBD-PC \cite{Avanti}.  To measure changes in the thickness of the glycerol layer, it is possible to simultaneously illuminate the fluid using an oblique laser line. Details about the experimental procedures are provided in \cite{Fallest-2010-FVS,Strickland2014}.  Sample images are shown in Figure~\ref{fig:exper}, with the laser visible in (a).

The simulations presented in this paper provide evidence that using an empirically-based equation of state improves agreement between solutions of the model and experimental observations.  In general, the dynamics of the surfactant profile morphology differ  above vs. below the critical surfactant concentration $\Gamma_c$. This effect has also been observed in outward spreading experiments, and suggests new inward spreading experiments. The simulations confirm that the shape of the fluid surface is only weakly dependent on the choice of equation of state, explaining why there has previously been reasonable agreement in the fluid profile between models and experiments even without an empirically-based equation of state. 

Importantly, the use of an empirically-based equation of state provides new insight into the  previous lack of agreement in the timescale of the simulations and experiments.  First, because the empirical equation of state has three distinct regions, a single timescale and spreading parameter may not be adequate to reproduce the timescale in experiments.  Second, although an initial fluid meniscus does not seem to have a major effect on dynamics, a surfactant meniscus could be created in experimental conditions and impact the timescale.  Third, horizontal shifts in the equations of state (including mis-identification of the critical monolayer concentration or the choice of an unrealistic model), can have significant impacts on both the spatial and temporal dynamics.  In contrast, the choice of nondimensional parameters seems unlikely to be the cause of lack of agreement.  The simulation results will therefore help guide future modeling efforts, as well motivate as new experimental explorations.

%=====================================================
\section{Mathematical Model} 

A set of equations first proposed by  \cite{Gaver-1992-DST} has often been used to model surfactant spreading on a thin viscous fluid film \cite{Fallest-2010-FVS,warner2004fingering, braun2012dynamics, peterson2011radial, witelski2006growing}.   The equation for the shape of the upper fluid surface (``fluid height'') $h(x,y,t)$ is based on the well-accepted thin film equation, which models the flow of a thin viscous fluid.  The height equation additionally incorporates the assumption that surfactant gradients induce surface stress through the tangential boundary condition.  The second equation for surfactant concentration $\Gamma(x,y,t)$ assumes that the fluid advects the surfactant by matching the velocities at the fluid-surfactant interface.  An ad-hoc term incorporates surfactant diffusion; an alternative derivation based on free-energy considerations has also been proposed \cite{pereira2007dynamics}. 

The resulting system of PDE in its common non-dimensional form is

\begin{equation}\label{e:film}
    h_t + \nabla\cdot\left(\frac{1}{2}h^2\nabla \sigma\right) =
    \beta\nabla\cdot\left(\frac{1}{3}h^3\nabla
    h\right)-\kappa\nabla\cdot\left(\frac{1}{3}h^3\nabla\nabla^2 h\right)
    \end{equation}
    
    \begin{equation}\label{e:surfactant}
    \Gamma_t + \nabla\cdot\left(h\Gamma\nabla \sigma\right) =
    \beta\nabla\cdot\left(\frac{1}{2}h^2\Gamma\nabla
    h\right)-\kappa\nabla\cdot\left(\frac{1}{2}h^2\Gamma\nabla\nabla^2
    h\right)+\delta\nabla^2\Gamma.
    \end{equation}
where $h(x,y,t)$ is the fluid height and $\Gamma(x,y,t)$ is the surfactant concentration. The gradient operator is two-dimensional ($\nabla = \partial_{x} \hat{x} +\partial_{y} \hat{y}$). We define $r=\sqrt{x^2+y^2}$ for use in some expressions below.   For detailed derivations of this well-studied model, please see \cite{levy2005partial, peterson2010flow}. 

\subsection{Non-dimensionalization paramters \label{sec:nondim-defs}}

The nondimensionalization in the above equations is standard:  $x=x_\mathrm{dim}/L$, $y=y_\mathrm{dim}/L$, $h=h_\mathrm{dim}/H$, and $\Gamma=\Gamma_\mathrm{dim}/\Gamma_c$, where $L,H$ are the lateral and vertical length scale, respectively, and $\Gamma_c$ is the critical monolayer concentration \cite{Kaganer1999,Reis2009}. Time is nondimensionalized as in \cite{Strickland2014}, motivated by \cite{gaver1990}: $t_\mathrm{dim} = \left( \frac{\mu L^2}{SH} \right) t $ where $\mu$ is the dynamic viscosity and $S \equiv \sigma_\mathrm{max}-\sigma_\mathrm{min}$ is the spreading parameter set by the max/min values of the surface tension. The three non-dimensional parameters in the model are 
$\beta \equiv \frac{\rho g H^2}{S}$
     (the ratio of gravity to capillary forces, based on fluid density $\rho$ and gravitational acceleration $g$),
$\kappa \equiv \frac{\sigma_\mathrm{max} H^2}{S L^2}$ 
     (the ratio of total to relative capillarity scaled by small parameter $H/L$), and
$\delta  \equiv \frac{\mu D}{SH}$
 (the inverse Peclet number, based on diffusion constant $D$). 

In the model equations, the $\nabla\sigma$ term  incorporates the effect of gradients in surfactant concentration through a constitutive relationship $\sigma(\Gamma)$ that relates surface tension $\sigma$ and surfactant concentration $\Gamma$.  The choice of a particular  equation of state (EoS) $\sigma(\Gamma)$ will be a major focus of this work. One important conclusion of our results, to be described in more detail below, is that modifying the time non-dimensionalization to include multiple timescales may be be necessary.

The simulations take the following values, consistent with 94\% pure glycerol at 20$^\circ$ \cite{glycvisc,glycdens}: 
glycerol density $\rho = 1.2$~g/cm$^3$, 
dynamic viscosity $\mu = 14$ poise, gravitational acceleration $g = 980$~cm/s$^2$, 
characteristic fluid depth $H=0.7$~cm, and
diffusion constant $D =  10^{−4}$~cm$^2$/s. 
In order to directly compare simulations of inward and outward spreading, we use a single characteristic lateral lengthscale $L= 3$~cm throughout this work.  This corresponds to a value similar to the dimension of the retaining ring in prior experiments \cite{Fallest-2010-FVS,Strickland2014,swanson2014surfactant}. 
Finally, the choice of maximum (clean glycerol) 
surface tension $\sigma_\mathrm{max} = 63.475$~dynes/cm, $\sigma_\mathrm{min} = 37.865$~dynes/cm which set the value of $S = 25.61$~dynes/cm will be made according to the discussion below (see \S\ref{sec:EEoS}).
Together, these choices set the non-dimensional model parameters  
$\beta = 2.44\times 10^{-1}$, $\kappa = 1.35 \times 10^{-3}$, and
$\delta = 7.81 \times 10^{-4}$ 
which we will refer to as the \emph{standard parameters}. 

%$\beta =(1.2 \cdot 9.8  \cdot 0.7^2)/(25.61) = 0.244$,
%$\kappa=(63.475  \cdot 0.7^2)/(25.61  \cdot 3^2)(1/100) = 0.00135$ and
%$\delta =(14 \cdot 0.0001)/(25.61 \cdot 0.7) \cdot 10 = 0.000781$. 

\subsection{Initial and Boundary Conditions}

The initial conditions are motivated by the laboratory experiments of \cite{Strickland2014} (see Figure~\ref{fig:exper}). We use several variations on a standard set of basic assumptions, all radially-symmetric. The standard fluid height initial condition for both inward and outward spreading simulations is uniform initial fluid height $h(r,0)=1$.  The standard surfactant concentration initial condition places a uniform layer of surfactant inside the retaining ring for outward spreading or outside the ring for inward spreading.  In both cases, surfactant spreading occurs towards regions with less surfactant, where the surface tension is higher. 

Inward spreading initial condition IC1 is used in Figures~\ref{fig:multiVStanh1}(a,c,e,g),\ref{fig:EEOS07}(a,c,e,g),\ref{fig:multiVStanh2}(a,c,e,g),\ref{fig:shiftEOSprofile}, and\ref{fig:beta},
\[
\begin{array}{cc}
 h(r,0)= 1.0  \hspace{.75in}
 &
\Gamma(r,0)= \begin{cases} 
      0, &  r  \leq 1 \\
      0.7  \, \mathrm{or} \, 2.0, &  r > 1  
   \end{cases}
\end{array}
\]
while for outward spreading IC2 is used in Figures~\ref{fig:multiVStanh1}(b,d,f,h),\ref{fig:EEOS07}(b,d,f,h),\ref{fig:multiVStanh2}(b,d,f,h),\ref{fig:outward_EoSshift}, and\ref{fig:outward_beta}  
\[
\begin{array}{cc}
 h(r,0)= 1.0  \hspace{.75in}
 &
\Gamma(r,0)= \begin{cases} 
      0.7 \,  \mathrm{or} \,  2.0, &  r  \leq 1 \\
      0, & r  > 1  
   \end{cases}
\end{array}
\]

In \S\ref{sec:pile}, we will modify these basic initial conditions to examine the effects of an annular-shaped surplus of fluid or surfactant in the vicinity of the retaining ring. This is motivated by the observation in experiments of a fluid/surfactant meniscus drawn up by the ring as it is slowly removed from the surface.  For spreading with additional fluid thickness (Figure~\ref{fig:fluidpile}), a piecewise constant initial condition simulates the presence of additional fluid at the ring location, while the surfactant initial conditions for inward and outward spreading remain the same as defined above. We use the additional (due to meniscus) fluid height $h_+$ as a parameter in IC3 (results shown in Figure~\ref{fig:fluidpile}):

\[
\begin{array}{cc}
h(r,0)= \begin{cases} 
      1.0, &  r  \leq 1 \\
      h_+, & 1 <  r  \leq \mbox{1.5} \\
      1.0, & \mbox{1.5} <  r  \leq \mbox{L} \\      
   \end{cases}
\end{array}
\]

For inward spreading with an additional surfactant in an annular region (see Figure~\ref{fig:addpile1}) we maintain a constant initial height of the fluid, and create a region of increased surfactant concentration $\Gamma_+$ which extends from $r=1$ out to a distance $1 + r_+$. This is initial condition IC4:
\[
\begin{array}{cc}
\Gamma(r,0)= \begin{cases} 
      0, &  r  \leq 1 \\
      \Gamma_+, & 1 <  r  \leq 1 + r_+ \\
      0.9, & 1+ r_+ <  r \leq \mbox{L} \\      
   \end{cases}
\end{array}
\]

Similarly, for outward spreading with an additional surfactant annulus in
Figure~\ref{fig:addpile3} we use IC5: 
\[
\begin{array}{cc}
\Gamma(r,0)= \begin{cases} 

      0.9, &  r  \leq 1 - r_+ \\
      \Gamma_+, & 1 - r_+ <  r \leq 1 \\
      0, & 1 <  r  \leq \mbox{L}  
   \end{cases}
\end{array}
\]

For all simulations we use the boundary conditions 
\begin{eqnarray}\label{e:boundary}
h_x=h_{xxx}=\Gamma_{x}=0,& x=-L \mbox{ and } x=L; \\
h_{y}=h_{yyy}=\Gamma_{y}=0,& y=-L \mbox{ and } y=L
\end{eqnarray}
and choose a spatial domain $[-L,L] \times [-L,L]$ for simulations large enough that the primary fluid and surfactant waves are away from the boundary.  There are $400$ gridcells for each $2\pi$ nondimensional units.  Grid refinement tests for this code were performed in \cite{conti2013effects}. The default value is $L= \pi$ and a larger domain $L=2\pi$ is used if waves approach the boundary of the smaller domain.  Note that this approach mimics that of the laboratory experiments, in which data are taken well away from the boundary formed by the walls of the cylindrical containment well.   
Our simulation code computes with a fully-2D discretization in $(x,y)$, to enables us to monitor the results for significant deviations from axisymmetry. Since none occur \cite{conti2013effects}, we plot only a single independent spatial dimension ($x$), for simplicity.

\subsection{Physically-Motivated Empirical Equation of State \label{sec:EEoS}}

As mentioned above, closing the system of equations requires an equation of state relating the surface tension $\sigma$ and the surfactant concentration $\Gamma$.  Previous models for the equation of state have been based on the fundamental premise that surface tension decreases as surfactant concentration increases. For simplicity, the earliest versions of the model \cite{Gaver-1992-DST} employed the linear equation of state (LEoS)

\begin{figure}
\centering
\includegraphics[height=2in]{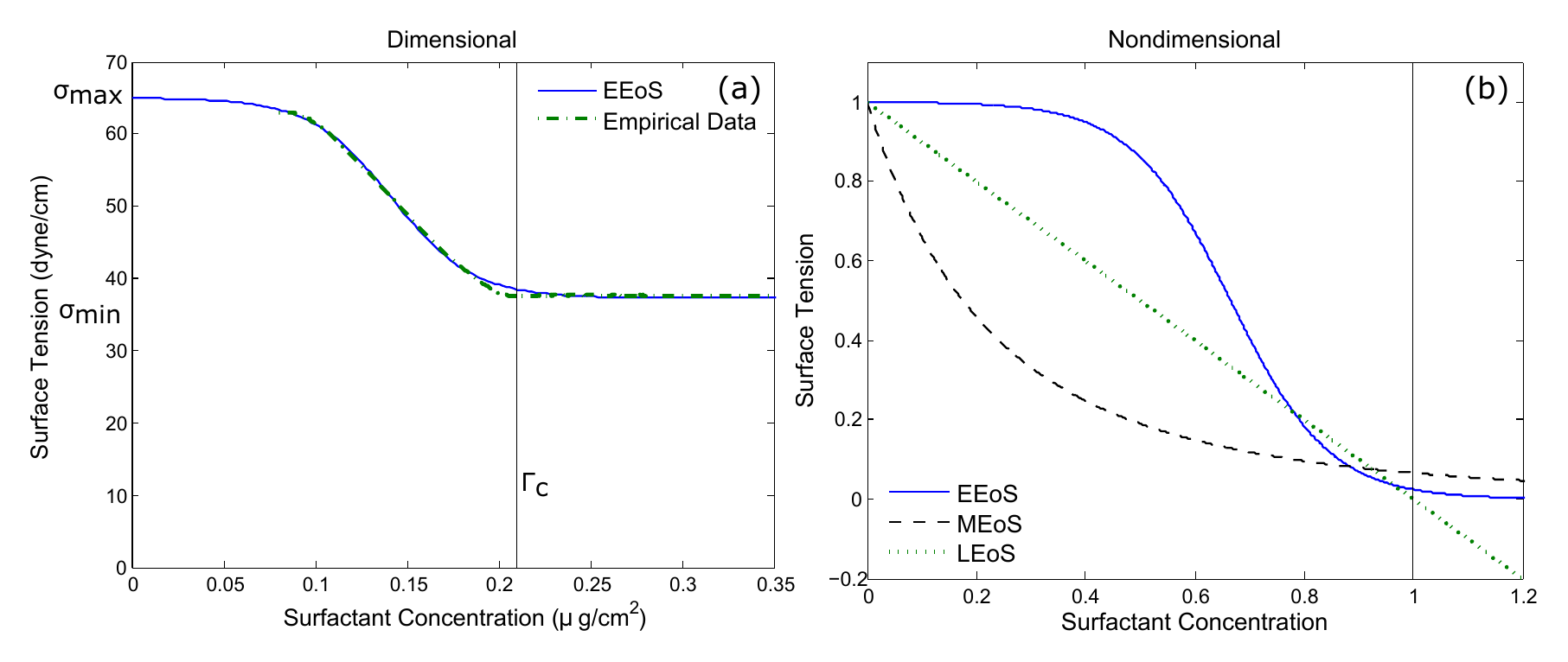}
\caption{(a) Empirical data and the fit curve relating surfactant concentration and surface tension \cite{strickland2015surfactant}. From this data, we obtain the experimental values $\tilde{\sigma}_\mathrm{min} = 37.416 \, \mathrm{dynes/cm}$,
$\tilde{\sigma}_\mathrm{max} = 63.026 \, \mathrm{dynes/cm}$,
and $\Gamma_c = 0.21 \, \mu\mathrm{g/cm}^2$. 
(b) Comparison of nondimensional linear LEoS from Eq.~\ref{eq:linear}, multilayer MEoS from Eq.~\ref{eq:multiEOS}, and empirical EEoS from Eq.~\ref{eq:tanhEOS}.
}
\label{fig:EOS}
\end{figure}

\begin{equation}
\sigma(\Gamma) = 1 -\Gamma.
\label{eq:linear}
\end{equation}  
While this LEoS has the advantage that the constant surfactant concentration gradient simplifies the analysis of the model equations,  the negative slope causes the surface tension to become negative at large surfactant concentrations.  To address this problem, a second model proposed in \cite{borgas1988monolayer} uses a multilayer equation of state (MEoS), which in its dimensional form 
has $\sigma(0) = \sigma_\mathrm{max}$
and decreases asymptotically to 
$\sigma(\Gamma) = \sigma_\mathrm{min}$ for large $\Gamma$. 
We use this MEoS in the nondimensional form 
\begin{equation}
\sigma(\Gamma) = (1+\eta \Gamma)^{-3},
\label{eq:multiEOS}
\end{equation}  
where $\eta = \sigma_\mathrm{max}/S$.    Note that neither multiplicative nor additive factors affect the simulations using the MEoS, since the former are removed by non-dimensionalization, and the later by taking the gradient.

In previous work \cite{Strickland2014,swanson2014surfactant}, we compared simulations using the LEoS and MEoS to data from spreading experiments.  This work demonstrated that neither EoS resolved disagreements between simulations and experiments in either timescale or spatial distribution of surfactant. In this work, we move beyond the ad-hoc linear and multilayer choices for EoS, choosing an equation of state which is consistent with empirical measurements, and simulating the effect of this new choice of $\sigma(\Gamma)$ on solutions of the mathematical model for a range of experimentally realistic intial conditions. Surprisingly, this approach has not yet been explored.  In addition, we examine the role played by the nondimensional groups $\beta$, $\kappa$, and $\gamma$ and test the sensitivity to initial conditions, as has long been done in studies of surfactants and thin liquid films \cite{craster2009dynamics,de1994nonlinear, oron1997long,gaver1990}.

To obtain empirical measurements of $\sigma(\Gamma)$, the standard technique is a   Langmuir-Blodgett trough (or Pockels scale). This apparatus measures the surface pressure while barriers compress the surfactant/lipid molecules located adsorbed to the surface of a liquid. For known container dimensions and a known quantity of surface molecules, the set of pressure and area measurements provide a plot $\sigma(\Gamma)$. (Note:  in the chemistry literature, the raw curve is often reported as the $\pi$-$A$ diagram directly relating the surface pressure $\pi$ to the area occupied by the molecular monolayer.)  Using this technique, it is possible draw on  empirical measurements to drive choice of a particular mathematical form of $\sigma(\Gamma)$ used in Eqns.~\ref{e:film},\ref{e:surfactant}.

To motivate a functional form for the empirical equation of state (EEoS), we examine data for a monolayer of NBD-PC on glycerol collected by Strickland \cite{strickland2015surfactant}, as plotted in Figure~\ref{fig:EOS}a.   A few important distinctions from the more commonly-used MEoS (Eq.~\ref{eq:multiEOS}) are worth noting. First, while the MEoS falls most steeply for low $\Gamma$ and has a single trend, the EEoS form of $\sigma(\Gamma)$ has three distinct regimes corresponding to low, intermediate, and high $\Gamma$. The surface tension falls most sharply for intermediate $\Gamma$, with the low and high $\Gamma$ values remaining approximately (but not precisely) constant. Because the gradient of $\sigma(\Gamma)$ appears in Eqns.~\ref{e:film},\ref{e:surfactant}, we will see that the slopes in all three regimes have a significant impact on the simulation results. 

Motivated by the experimental results, we consider a new model EoS which can capture all three regimes.  While a piecewise linear function with three regions would be analytically-convenient (constant gradient), and capture primary features of the data except for the curvature, it has the serious disadvantage of having discontinuous derivatives. In addition, the slope of the EoS should never be zero, as the model would then predict a non-physical pile-up of fluid.  A negative slope would cause the same issue for large values of $\Gamma$ as the LEoS.   In considering the shape of Figure~\ref{fig:EOS}a, we find that a hyperbolic tangent function
\begin{equation}
 \sigma(\Gamma) =  \frac{S}{2}\tanh(k_1(\Gamma - k_2)) + k_3
 \label{eq:tanhEOS}
 \end{equation}
models the empirical data as our EEoS while also remaining continuous and differentiable.  
We used the experimental data $\tilde{\sigma}_\mathrm{min} = 37.416$ and $\tilde{\sigma}_\mathrm{max} = 63.026$ from Figure~\ref{fig:EOS}a in an two-step process to obtain an empirical value of $S = 25.610$.  First, an optimization routine in Matlab provided the fitted values $k_1 = -26.31$, $k_2 = 0.14$ and $k_3 = 50.67$ in equation~\ref{eq:tanhEOS}.  Second, we redefine the values of $\sigma_\mathrm{max} = k_3 + S/2 = 63.475$ and $\sigma_\mathrm{min} = k_3 - S/2 = 37.865$ so that they correspond to the fitted curve that will be used in the simulations.

We nondimensionalize the EEoS using $\Gamma_\mathrm{nondim}=\Gamma_\mathrm{dim}/\Gamma_c$ and $\sigma_\mathrm{nondim}=(\sigma_\mathrm{dim}-\sigma_\mathrm{min})/S$. The critical surfactant concentration $\Gamma_c$ is the value of $\Gamma$ at which the trough data indicate that additional surfactant does not reduce the surface tension. This value is obtained by finding the local minimum in the data at $\Gamma_c=0.21$ in the EEoS (see vertical lines in Figure~\ref{fig:EOS}). This is a lower value than previous papers, which used $0.3$ as an approximation of $\Gamma_c$.  To provide a consistent comparison, we will use a MEoS (see equation \ref{eq:multiEOS}) derived from the same values for $\sigma_\mathrm{max}$, $\sigma_\mathrm{min}$, $\Gamma_c$  and $S$, with $\eta \equiv \sigma_\mathrm{min} / (\sigma_\mathrm{max} - \sigma_\mathrm{min}) = 1.48$.  Figure~\ref{fig:EOS}b  has all three curves, the LEoS, MEoS and EEoS.

The simulations of this paper are performed using an open-source code described in \cite{ClaridgeGitHubPaper}, with code and documentation freely available on Github \cite{ClaridgeGitHub}.  In previous work by this group and others, model equations ~(\ref{e:film},\ref{e:surfactant}) have been solved using many approaches \cite{witelski2006growing, warner2004fingering,Strickland2014}; the advantage of our code is that it facilitates easy modification of terms in the equation and boundary conditions and provides a package for convergence testing.  The second-order scheme is based on a finite volume approach (using Newton's Method and BiCGStab), which takes advantage of the free open source Clawpack package and enables the user to compute solutions with small (or zero) coefficients on the regularizing terms.  

%=================================================================
\section{Results \label{sec:results}}

We present numerical solutions to the  system of PDE~(\ref{e:film},\ref{e:surfactant}) using an empirically-derived EoS, realistic initial and boundary conditions, and appropriate model parameters $\beta, \kappa, \delta$.  We frame our investigations as answers to four key questions:

\begin{enumerate}
\item \S\ref{sec:EEos} {\itshape Choice of EoS shape:} How do the general dynamics of solutions for the multilayer EoS (MEoS) differ from those for the empirical EoS (EEoS)? 
\item \S\ref{sec:shift} {\itshape Offsets to the EoS:} What is the effect of shifting the EEoS vertically (offset in $\sigma$) or horizontally (offset in $\Gamma$)? 
\item \S\ref{sec:nondim} {\itshape Dependence on non-dimensional parameters:} What is the effect on simulations with the EEoS of varying the non-dimensional parameters? 
\item \S\ref{sec:pile} {\itshape Effect of retaining ring:} Do simulations with the EEoS indicate that the retaining ring that creates the initial surfactant distribution has a strong effect on the spreading dynamics?   
\end{enumerate}

\noindent We answer these questions by focusing on two features also observable in experiments:  the fluid height $h_c(t) \equiv h(0,t)$ at the center of the domain (which is affected by the capillary ridge) and the location $r_s(t)$ of the leading edge of the surfactant as it spreads.  In addition, we discuss the implications of these results for future laboratory experiments.

In viewing simulation results, keep in mind that they are computed in two spatial dimensions $(x,y)$, but plotted as $h(x,t)$ and $\Gamma(x,t)$.  As noted above, for symmetric initial conditions such as ours, deviations from axisymmetry are not significant  \cite{conti2013effects}.  Also note that because the size of surfactant molecules is insignificant compared to that of the fluid depth, the model assumes the surfactant does not to add to the height in the fluid/surfactant system. Thus, the $\Gamma(x,t)$ plots represent the local surface concentration across the diameter of the well; physically, this corresponds to a  more-densely or less-densely packed layer.

All simulations are run with the standard parameters unless otherwise noted.  The solutions to the surfactant equation do not have compact support due to diffusion ($\beta>0$ in the model as well as reality). Therefore,  we must choose an effective location of the leading surfactant front, which we define as the location where $\Gamma = 0.01$.

%=========================================================
\subsection{\textbf{Investigation 1:  Choice of EoS shape} \label{sec:EEos}}

%\begin{SCfigure}

\begin{figure}
\includegraphics[width=4.7in]{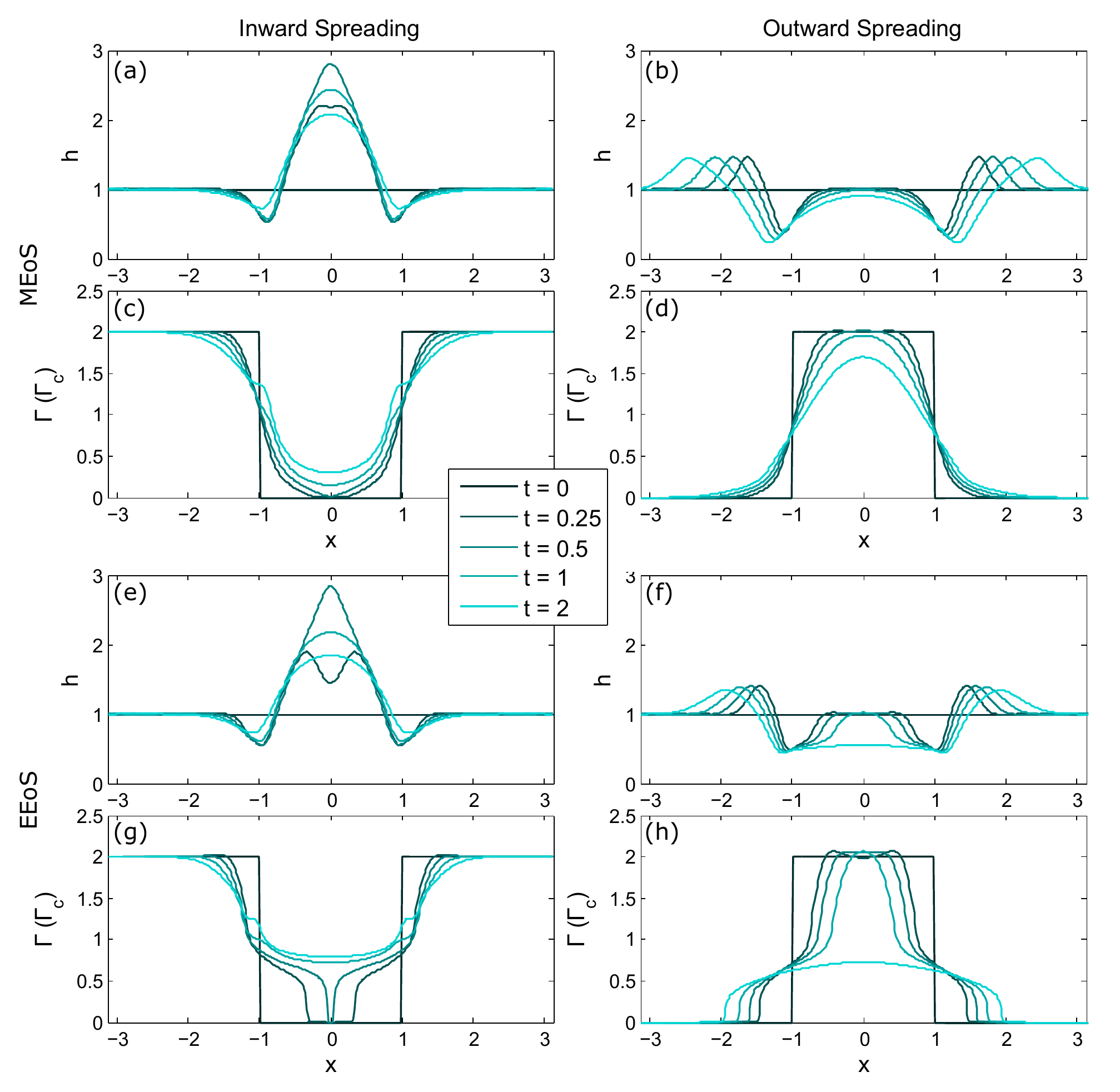}
\caption{Typical spreading dynamics for each equation of state with standard parameters and initial surfactant concentration $\Gamma=2.0>\Gamma_c$).  The upper quartet of plots (a,b,c,d) use the MEoS whereas the lower quartet (e,f,g,h) use the EEoS.   Plots (a,b,e,f) are fluid profile dynamics $h(x,t)$ and (c,d,g,h) are surfactant concentration profiles $\Gamma(x,t)$.  Left plots (a,c,e,g) have an inward spreading initial condition (IC1) and right plots (b,d,f,h) have an outward spreading initial condition (IC2).
}
\label{fig:multiVStanh1}
\end{figure}
%\end{SCfigure}

\begin{figure}
\includegraphics[width=4.7in]
{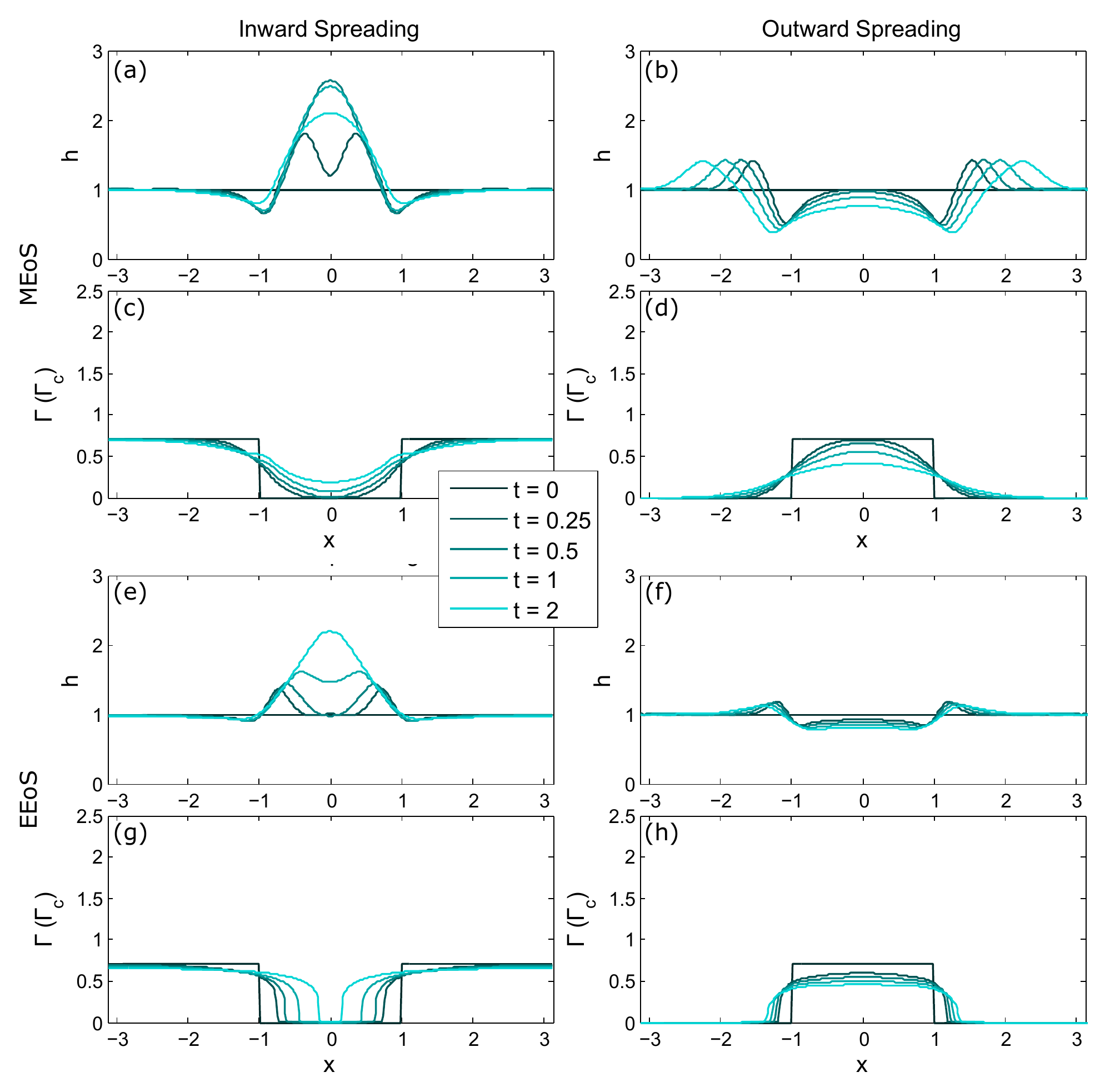}
\caption{Typical spreading dynamics for each equation of state with standard parameters and initial surfactant concentration $\Gamma=0.7<\Gamma_c$). The upper quartet of plots (a,b,c,d) use the MEoS whereas the lower quartet (e,f,g,h) use the EEoS.   Plots (a,b,e,f) are fluid profile dynamics $h(x,t)$ and (c,d,g,h) are surfactant concentration profiles $\Gamma(x,t)$.   Left plots (a,c,e,g) have an inward spreading initial condition (IC1) and right plots (b,d,f,h) have an outward spreading initial condition (IC2).  
} 
\label{fig:EEOS07}
\end{figure}

We begin by answering the most general question:  what are the most significant effects of switching to the empirical equation of state (EEoS) in place of the more commonly used multilayer equation of state (MEoS)?  Figures~\ref{fig:multiVStanh1} and \ref{fig:EEOS07} show typical results at above and below the critical monolayer concentration, respectively. In both figures, we present solutions for  inward (a,c,e,g) and outward (b,d,f,h) surfactant spreading.  The top quartet of plots (a-d) presents simulations using the MEoS and the bottom quartet (e-h) presents simulations using the EEoS. Each profile is a snapshot in time, from darkest at $t=0$ to lightest at $t=2$. 

\subsubsection{Inward spreading}

During inward-spreading, the fluid develops an inward-moving annular capillary ridge as it is pulled by the surfactant spreading into the central (clean) region (Figures~\ref{fig:multiVStanh1}a,c,e,g and \ref{fig:EEOS07}a,c,e,g). The fluid ridge coalesces into a single central maximum and then relaxes to an equilibrium at the original uniform height ($h(x,t)=1$). These general dynamics of fluid coalescence, central growth and decay were previously observed in  laboratory experiments using laser profilometry \cite{Fallest-2010-FVS,Strickland2014,swanson2014surfactant}, and are similar for either choice of EoS. However, there is an important distinction in $h(x,t)$: for simulations run with the MEoS (Figure~\ref{fig:multiVStanh1}a,b  and \ref{fig:EEOS07}a,b) the annular fluid capillary ridge (double-peaked structure) coalesces more quickly than those with the EEoS. This result explains why in \cite{Strickland2014} there was surprisingly good morphological agreement in the experiment and simulation fluid profiles with the MEoS given the poor agreement in timescale and surfactant distribution.

The surfactant concentration profiles $\Gamma(x,t)$ are more distinct for the two EoS choices. For surfactant layers with an initial condition above $\Gamma_c$ (Figure~\ref{fig:multiVStanh1}), the  EEoS surfactant profile has leading ``foot'' that pushes into the central region, with a pronounced and steep leading edge. In contrast, the MEoS surfactant curve retains a smoother profile at all times, within only a small kink near the location of the ring (later associated  with the dip in the fluid at that location).  (Note that this is different than the precursor ``foot" observed in \cite{tiberg1994spreading}, which was a result of interaction between surfactant and a solid substrate, not surfactant and thin liquid film.)  Even for surfactant layers with an initial condition below $\Gamma_c$ (Figure~\ref{fig:EEOS07}), the EEoS is better able to maintain strong gradients than the MEoS. Experiments for inward spreading have not been performed with $\Gamma > \Gamma_c$; this is a prediction that could be tested in future experiments.  In the figures that follow, we will focus on the parameters used in Figure~\ref{fig:multiVStanh1} (initial conditions above a monolayer) since the surface flow in this case is more sensitive to the choice of EoS due to the simultaneous presence of regions of high, middle, and low surfactant concentrations. 

\subsubsection{Outward spreading}

During outward-spreading, the fluid develops an outward-moving annular capillary ridge as it is pushed by the surfactant spreading into the outer (clean) region (Figures~\ref{fig:multiVStanh1}b,d,f,g and \ref{fig:EEOS07}b,d,f,g). This feature is present independent of the choice of EoS, and for initial surfactant concentrations above and below $\Gamma_c$.  However, for the MEoS simulations, the height $h_c$ of the central peak decays slowly even for simulations run much longer, whereas with the EEoS, there is a central fluid depression that extends across the entire region in which the surfactant had initially been deposited.  The EEoS behavior is consistent with what is observed in experiments \cite{swanson2014surfactant}, and thus is a better model.

As with inward spreading, the surfactant concentration profiles $\Gamma(x,t)$ are even more distinct for the two EoS choices.  Again the surfactant layers with an initial condition above $\Gamma_c$ produce a foot-like layer that emerges from the central region, ending in a pronounced leading edge. The sharp decrease in surfactant concentration coincides with the location of the fluid  capillary ridge. Below $\Gamma_c$ the foot layer extends with no reservoir.  As shown in Figure~\ref{fig:exper}bc, these same morphologies are present in the experiments.  In the experimental (top view) images  above $\Gamma_c$ a bright central region is surrounded by a lower-intensity foot behind the leading edge.  In the images  below $\Gamma_c$ no reservoir is present.  These important features are not reproduced by the MEoS-based simulations.  Instead, the MEoS case shows a consistent shape as the initial central surfactant layer decays and the leading edge seems to show little evolution. Therefore, the reservoir and foot-like features provide a striking improvement in morphological agreement between experiment and simulation by using the EEoS as compared to the MEoS. 

In all cases (inward/outward, MEoS/EEoS), the fluid profile maintains its initial depression at the surfactant boundary in the initial condition (corresponding to the retaining ring location in the experiment). In Figure \ref{fig:multiVStanh1} the outward spreading (right) central height evolution is much more distinctive between choices of EoS than in the inward spreading simulations.  The MEoS has a smooth decay over time, while the EEoS has a growth phase and a steep decay phase.  The surfactant leading edge plots also are more distinct; they have similar shapes, but the leading edge with the EEoS has a lower velocity.

\subsubsection{Timescale}

\begin{figure}
\centering
\includegraphics[height=4.5in]{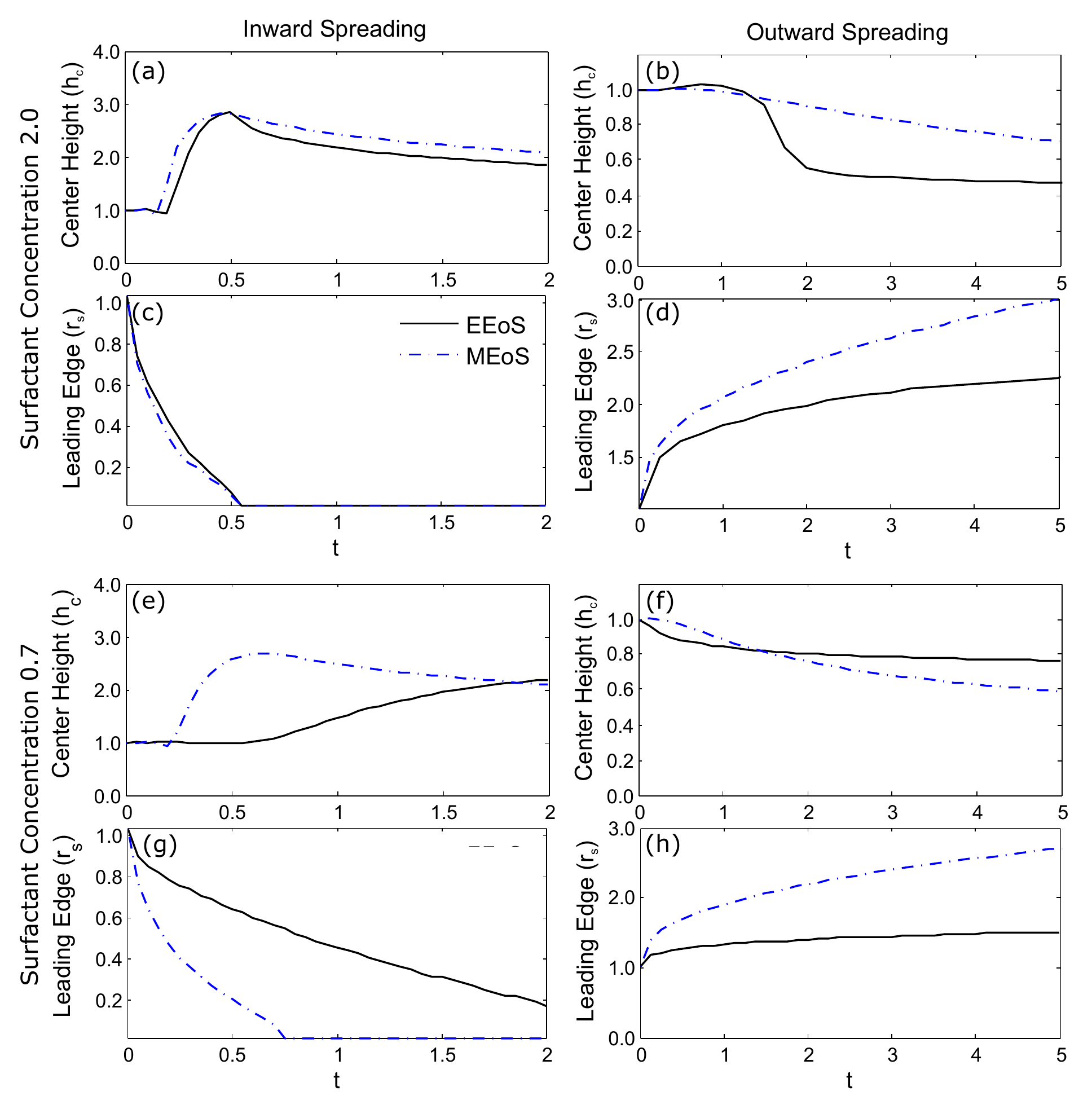}
\caption{Comparison of $h_c(t)$ (a,b,e,f) and $r_c(t)$ (c,d,g,h) for simulations of Figure~\ref{fig:multiVStanh1} (a,b,c,d) and Figure~\ref{fig:EEOS07} (e,f,g,h) using IC1 and standard parameters.  
}
\label{fig:multiVStanh2}
\end{figure}

Figure~\ref{fig:multiVStanh2} illustrates how the key dynamics from Figs.~\ref{fig:multiVStanh1} and \ref{fig:EEOS07} can be captured by considering only the height of the central peak ($h_c$) and the location of the leading edge of surfactant ($r_s$). These plots compare the effect of MEoS and EEoS on these two dynamics. Importantly, they need not agree. This suggests that a second timescale, beyond  $t_\mathrm{dim}=(\frac{\mu L^2}{SH})t$, the one used in the non-dimensionalization.
This situation arises because only the  gradient of the equation of state appears in the system of PDE (\ref{e:film},\ref{e:surfactant}). In the LEoS this gradient is negative and constant, in the MEoS the gradient is negative and gradually decreasing, and in the  EEoS there are three distinct regions. At low and high $\Gamma$, the gradient is negative and small, but for intermediate surfactant concentrations the gradient changes dramatically in magnitude. It is therefore unlikely that a single parameter $S$ captures the magnitude of the gradient, and therefore there is no single timescale. This observation may explain why, in all previous comparisons of simulations and experiments, it has been necessary to re-dimensionalize the simulations using a different (shorter by a factor of 2 to 10) timescale than the model \S\ref{sec:nondim-defs} would predict  \cite{swanson2014surfactant,strickland2015surfactant}.

%====================================================================================
\subsection{\textbf{Investigation 2: Offsets to the EoS }\label{sec:shift}}

The choice of a particular lipid will determine a unique EoS, specific to that lipid \cite{Kaganer1999,Reis2009}. However, the general shape shown in Figure~\ref{fig:EOS} exhibits many features common to a number of lipids. Therefore, it is important to ask what features of the solutions change when shifting  the EEoS vertically (offset in $\sigma$) or horizontally (offset in $\Gamma$). To exemplify a few basic behaviors, we perform simulations in which we  shift the EEoS 30\% each direction: to the left $\sigma(\Gamma + 0.3)$, right  $\sigma(\Gamma - 0.3)$, up  $\sigma(\Gamma) + 0.3$ and down $\sigma(\Gamma)-0.3$. Because only the gradient of the EoS appears in the model (as seen in equations (\ref{e:film},\ref{e:surfactant}), vertical shifts should not affect the results, which is confirmed in our simulations.  Understanding the effects of horizontal shifts will allow us to test which features of the chosen EoS are essential for making quantitative comparisons with experiments.

\begin{figure}
\centering
\includegraphics[height=2.2in]{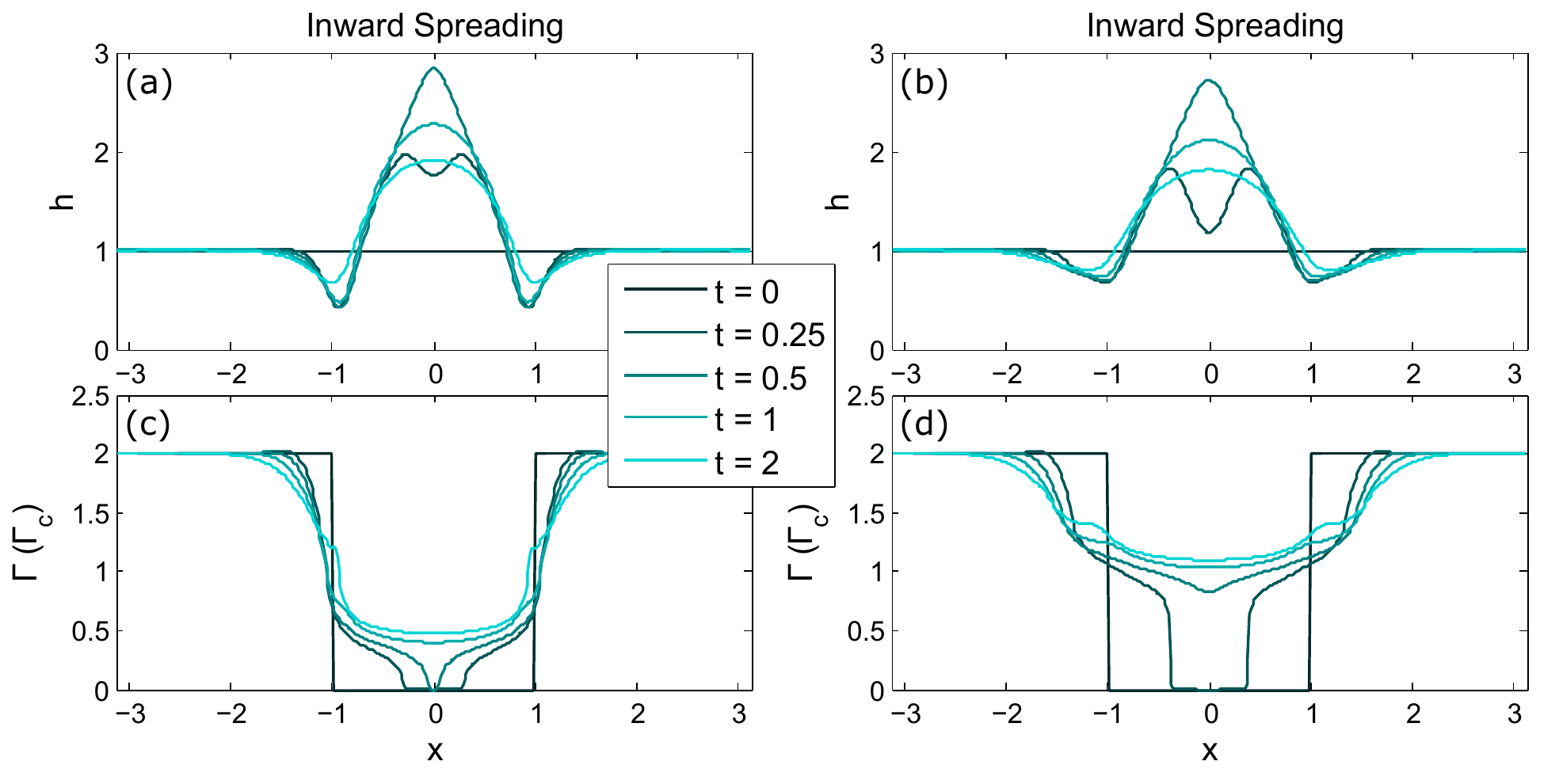}
\includegraphics[height=2.2in]{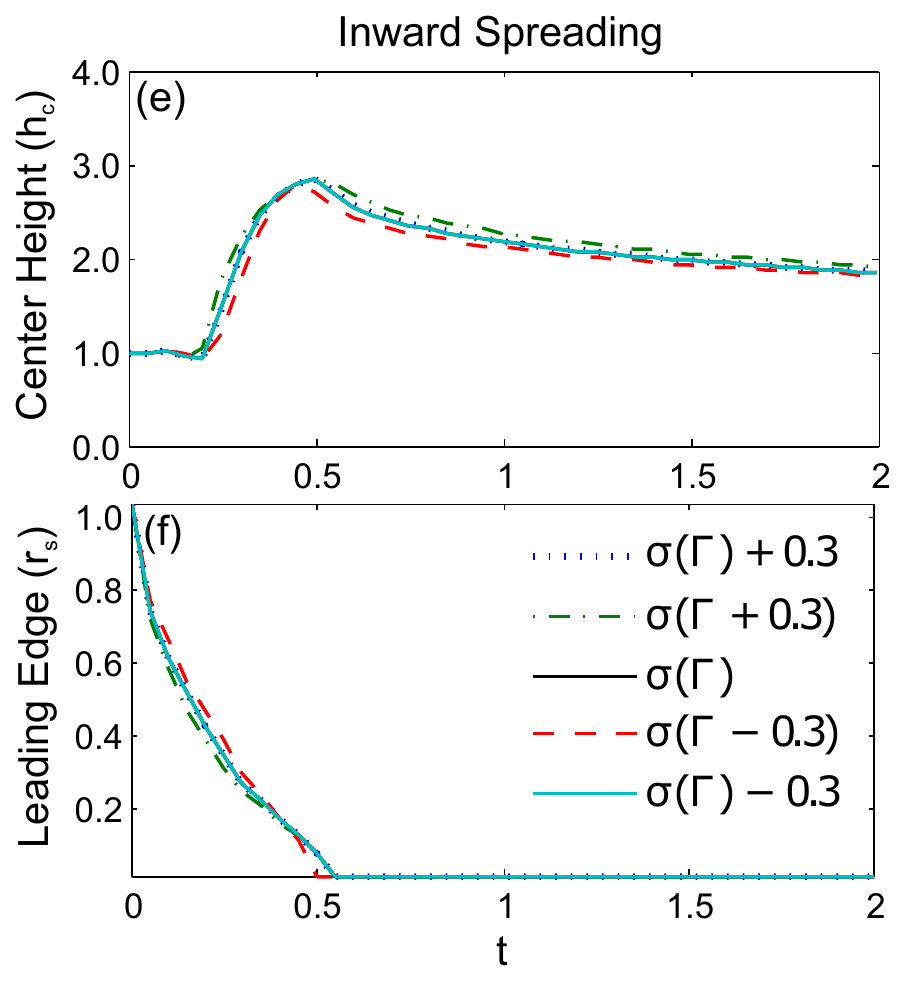}
\caption{Dynamics  for standard parameters and initial condition (IC1) using left shifted EEoS  $\sigma(\Gamma + 0.3)$ in plots (a,c) and right shifted EEoS $\sigma(\Gamma - 0.3)$ in plots (b,d). Characterizing (e) $h_c(t)$ and (f) $r_s(t)$ for four representative shifts in the EEoS.  
}
 \label{fig:shiftEOSprofile}
 \end{figure}

\medskip
\noindent {\itshape Inward spreading:} As shown in Figure~\ref{fig:shiftEOSprofile}e,  the simulations with the left-shifted EEoS $\sigma(\Gamma+0.3)$ have a higher minimum value at the center at early times.  The simulations with this EEoS shift also make the dynamics faster with peak in $h_c(t)$ and closure in $r_s(t)$ occurring at earlier time in Figures~\ref{fig:shiftEOSprofile}e,f. In the time snapshots of Figures~\ref{fig:shiftEOSprofile}a,b the left shift also causes earlier coalescence of the annular capillary fluid ridge.  In the surfactant profiles $\Gamma(x,t)$ of Figures~\ref{fig:shiftEOSprofile}c,d, the left shifted EEoS creates a smaller concentration of about $0.25$ in the foot, whereas with the right shift $\sigma(\Gamma-0.3)$, the foot concentration is about $0.7$.  These features are observable in experiments.  As expected, there is no difference for vertical shifting  and small differences for horizontal shifting in Figure~\ref{fig:shiftEOSprofile}e,f.

\begin{figure}
\centering
\includegraphics[height=2.5in]{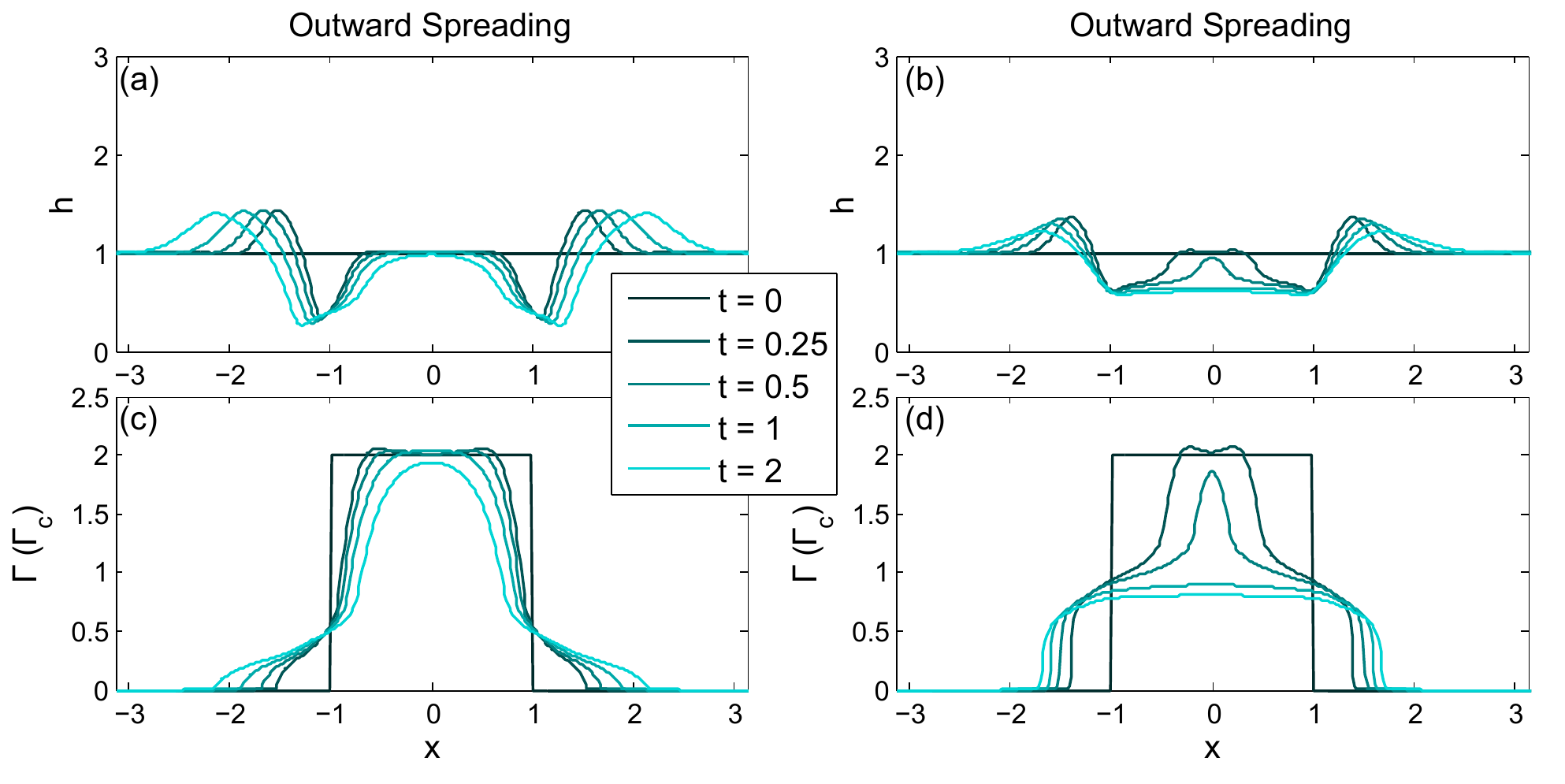}
\includegraphics[height=2.5in]
{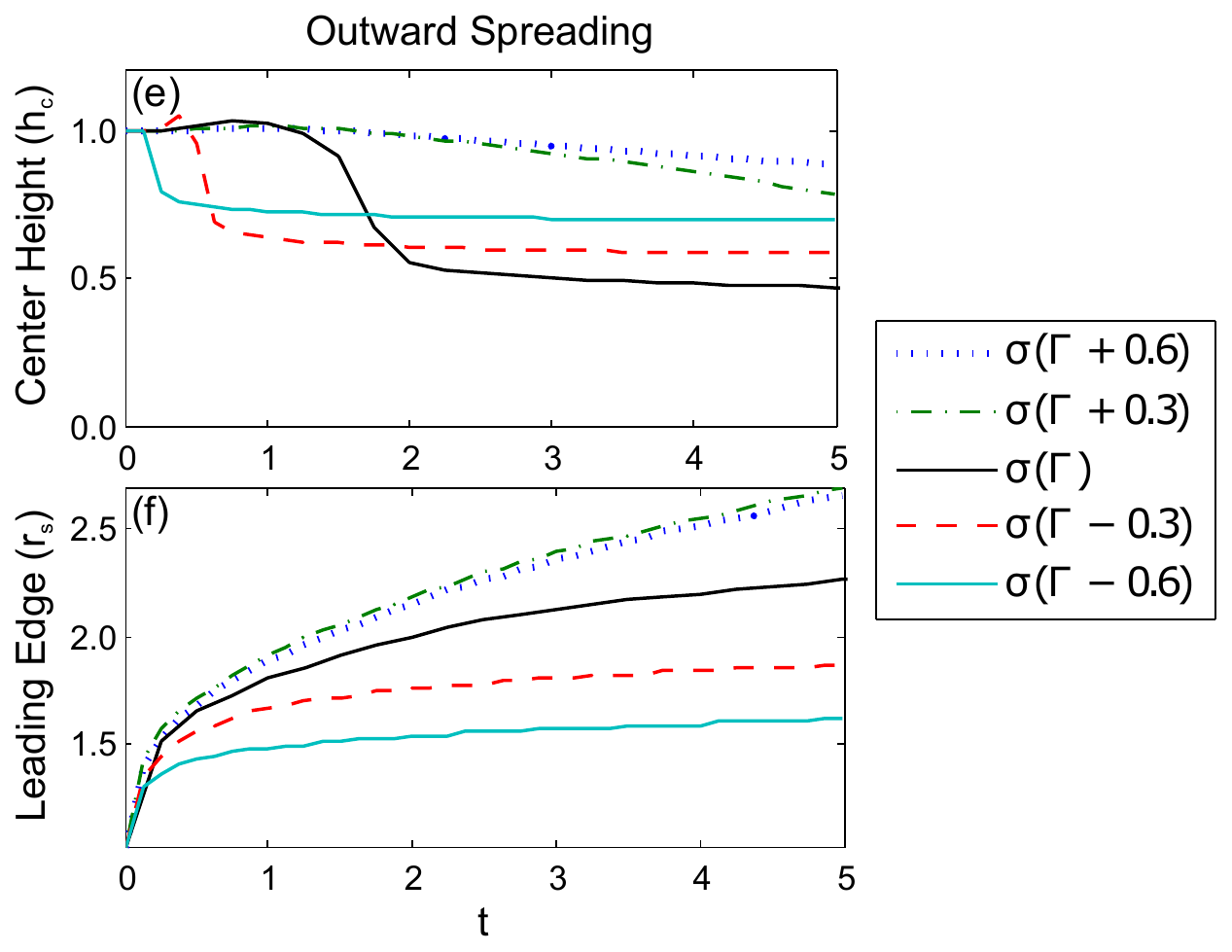}
\caption{Outward spreading sample dynamics  for standard parameters and initial condition (IC2) using left shifted EEoS  $\sigma(\Gamma + 0.3)$ in plots (a,c) and right shifted EEoS $\sigma(\Gamma - 0.3)$ in plots (b,d). 
(e,f) Characterizing (e) $h_c$ and (f) $r_s$ for the EEoS $\sigma(\Gamma \pm 0.3)$ and $\sigma(\Gamma \pm 0.6)$ with standard parameters.}
\label{fig:outward_EoSshift}
\end{figure}

\medskip
\noindent {\itshape Outward Spreading:} We observe that outward spreading is much more sensitive to horizontal offsets to the EoS than inward spreading, as shown in  Figure~\ref{fig:outward_EoSshift}. Here, we additionally include larger (60\%) shifts in the location of $\Gamma_c$, and do not consider vertical shifts because they have no effect.  We observe that shifting the EEoS to the left, $\sigma(\Gamma+0.3)$ and $\sigma(\Gamma+0.6)$, forms solutions that look much like the central height and leading edge results for the MEoS of Figure~\ref{fig:multiVStanh2}.  In both cases the central height $h_c(t)$ has a consistently small gradient and the surfactant leading edge $r_s$(t) advances more rapidly than for the original $\sigma(\Gamma)$.  Note that this similarity occurs because shifting the EEoS to the left shifts the region of the function with steeper gradient to low surfactant concentrations (such as those near $r_s$(t), which makes its gradient more like that of the MEoS.    

Shifting to the right, $\sigma(\Gamma-0.3)$ and $\sigma(\Gamma-0.6)$, moves the steeper portion of the EoS to larger surfactant concentrations.  For our initial conditions this accelerates the fluid growth and decay phases so that  $h_c(t)$ is quite steep, especially in the decay phase.  However, the leading edge of the surfactant $r_s(t)$ (where the surfactant concentration $\Gamma$ is lower) advances more slowly, since the large gradient in the EoS has been shifted to larger surfactant concentrations.

%============================================================================
\subsection{\textbf{Investigation 3:  Dependence on non-dimensional parameters} \label{sec:nondim}}

The values for the non-dimensional parameters ($\beta, \kappa, \delta$) in the model are derived from fluid and physical properties of the system, and in general cannot be independently varied in an experiments. Simulations provide a means to test the effects of each. We performed simulations for one-quarter, half and double 
the  standard parameter values for $\beta$, $\delta$ and $\kappa$, and found that the only notable changes occurred as a function of $\beta$, which we will explore here for both inward and outward spreading.  These effects would be difficult to detect in experiments. Because $\beta \equiv \frac{\rho g H^2}{S}$, the only way to change $\beta$ without changing the other parameters is to choose a fluid of a different density $\rho$, but this choice of a new material would also change $S$. 

\begin{figure}
\centering
\includegraphics[height=2.5in]{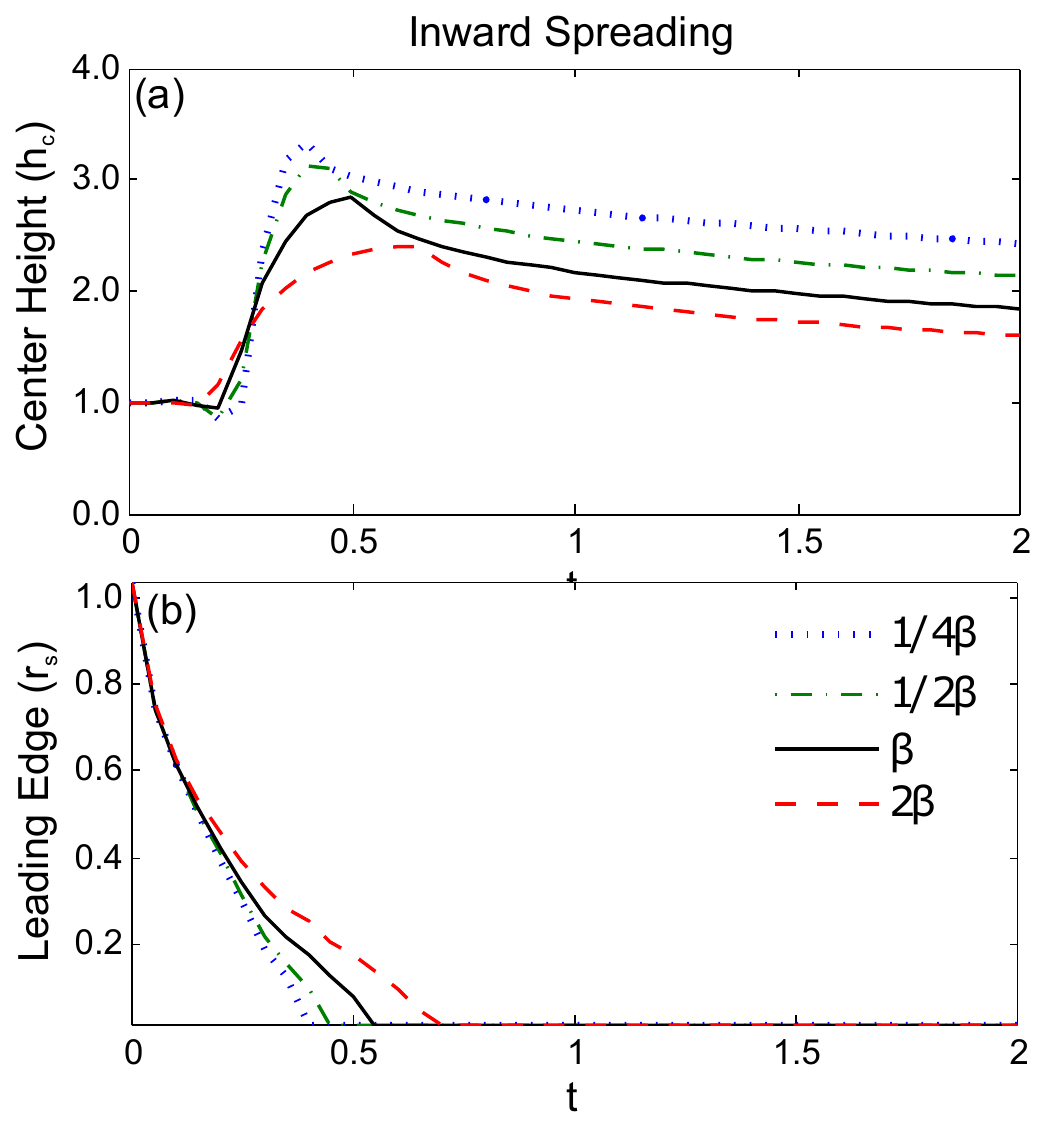} \\
\includegraphics[height=1.9in]{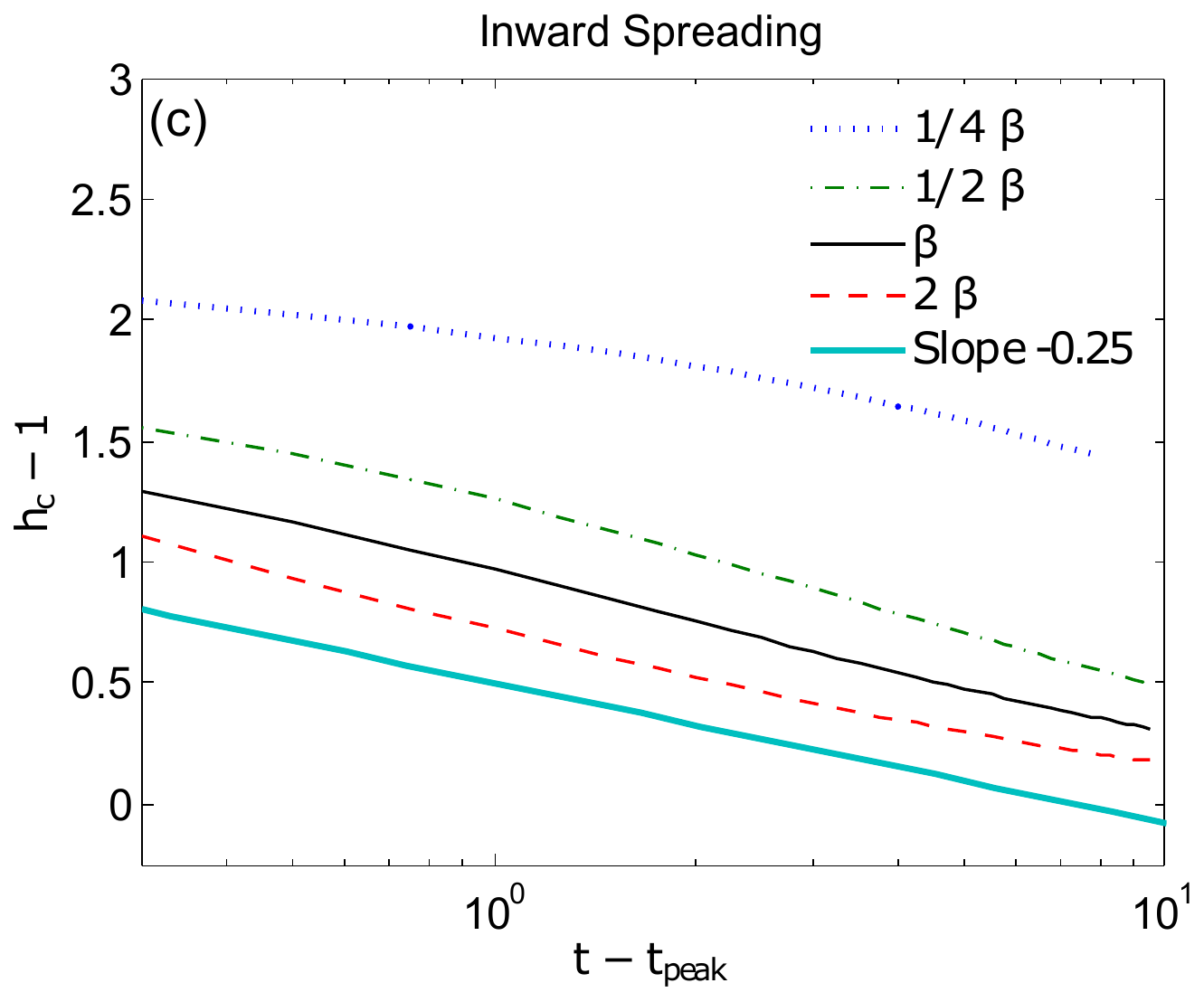}
\includegraphics[height=1.9in]{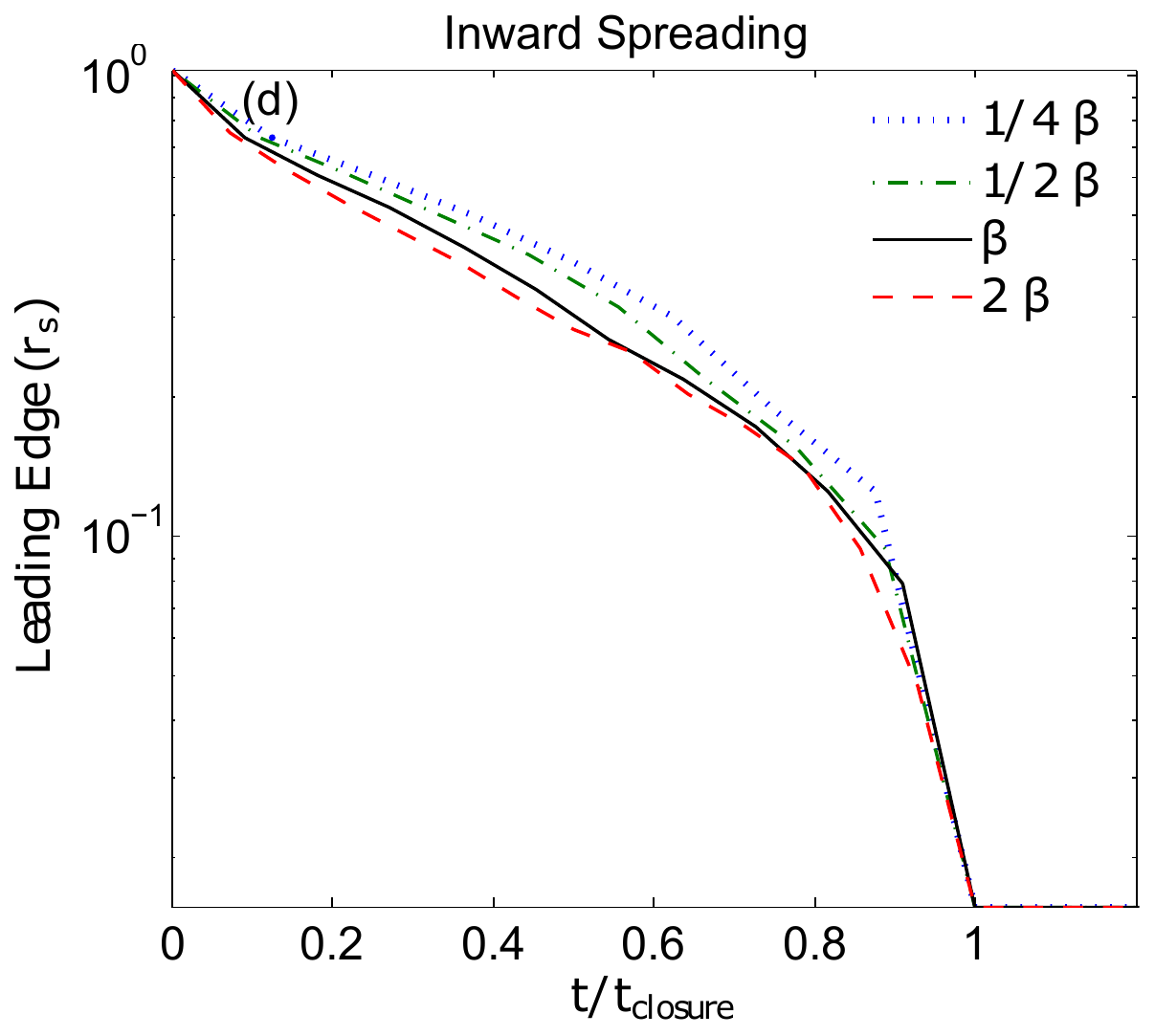}
\caption{Inward spreading simulations at fixed initial conditions (IC2), varying the relative importance of gravitational and capillary forces, through the non-dimensional parameter $\beta$.   Plots of (a) central fluid height $h_c(t)$ and (b) surfactant leading edge location $r_s(t)$.  Self-similar behavior of the decay of the fluid at center (c) and motion of surfactant leading edge (d). The time $t_\mathrm{peak}$ is defined as the time when $h_c(t)$ is at its maximum value and closure time is defined as the time when the leading edge location is $r_s=0$. The cyan (lowest solid) line in (c) is given by $h_c-1=-0.25 \ln(t-t_\mathrm{peak})+0.5$. 
}
\label{fig:beta}
\end{figure}

\medskip
\noindent{\itshape Inward spreading:} As shown in Figure~\ref{fig:beta}a,b the dynamics of $h_c$ and $r_s$ are qualitatively similar, independent of $\beta$. There are some small differences for a particular choice of $\beta$: smaller values (lower gravity, density or $H$ compared to spreading parameter $S$) can produce a larger fluid peak at the center, more rapidly.  We can quantify the similarity by considering the relaxation from a central peak at time $t_\mathrm{peak}$ back to a uniform fluid height $h(x,t) =1$. We observe that these dynamics follow a logarithmic decay (see Figure~\ref{fig:beta}c). If we take the time for the surfactant to reach $r=0$ to be a characteristic closure time, then the $r_s(t)$ dynamics are also independent of the choice of $\beta$.

\begin{figure}
\includegraphics[width=2.5in]{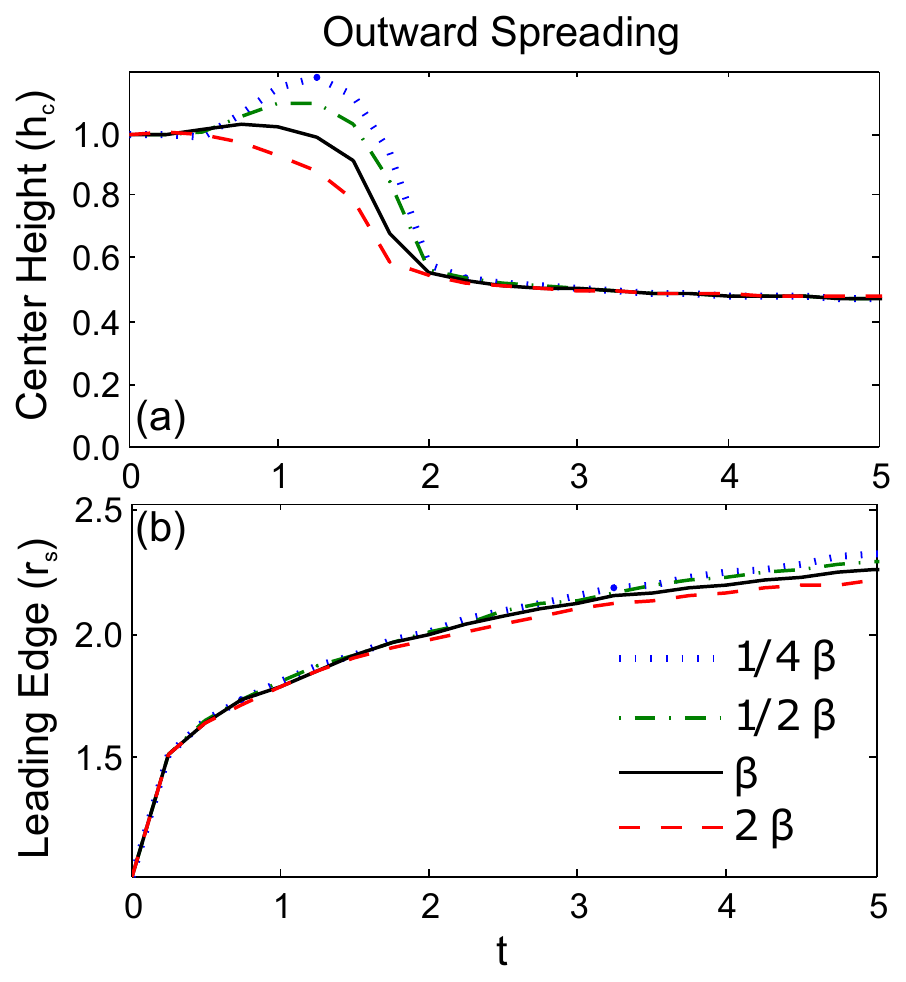}
\caption{Outward spreading simulations with nonstandard parameters to show the effect of varying $\beta$ on outward spreading using initial condition (IC3).  Plots of (a) central fluid height $h_c(t)$ and (b) surfactant leading edge location $r_s(t)$.}
\label{fig:outward_beta}
\end{figure}

\medskip\noindent {\itshape Outward spreading:} Similar effects are observed for outward-spreading in Figure~\ref{fig:outward_beta}, with smaller $\beta$ resulting in a taller capillary ridge, but with a largely invariant timescale (compare Figure~\ref{fig:outward_beta} to Figure~\ref{fig:beta} where the timescale varies).  A slight trend of faster spreading for lower $\beta$ is also present.

%====================================================================================
\subsection{\textbf{Investigation 4: Effect of retaining ring} \label{sec:pile}}

The first three investigations focus on solutions of the mathematical model, which has been compared to experiments.  This final investigation uses simulations to investigate the impact of  the experimental apparatus.  A key difference between experimental and numerical investigations is the necessity of using a retaining ring to set up the initial conditions when performing experiments. When this ring is lifted, a meniscus forms and then releases back to the surface after pinch-off. This raises the possibility that additional fluid or surfactant is present at the original location of the retaining ring. Below, we examine the effect of this additional material on the dynamics of inward/outward spreading. Initial conditions (IC3-5) are chosen to mimic conditions that could occur experimentally, and determine whether these conditions could have a significant impact on the timescale of the dynamics, as well as the spatial distribution of the surfactant.

\subsubsection{Fluid annulus}

\begin{figure}
\centering
\includegraphics[height=2.5in]{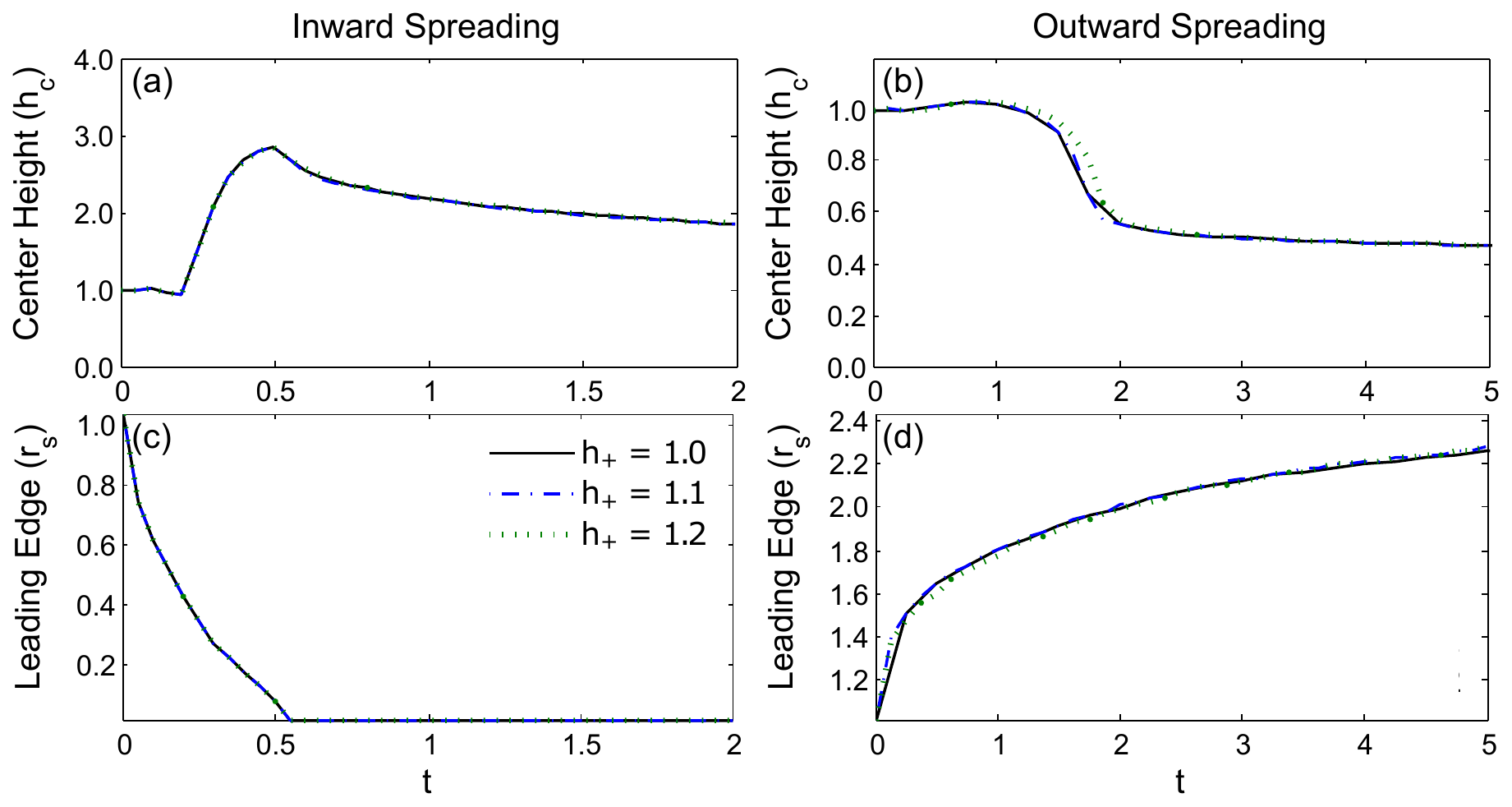}
 \caption{Inward and outward spreading simulations with standard parameters and initial condition (IC3)  comparing central fluid height $h_c(t)$ (a,b) and surfactant leading edge $r_s(t)$ (c,d) for no fluid annulus and with an annulus 10\% and 20\% above the original level.}
\label{fig:fluidpile}
 \end{figure}

We first consider the case of starting from IC which place an additional  annulus of fluid at the location of the ring, to mimic the after-effects of meniscus pinch-off. As shown in Figure~\ref{fig:fluidpile}, even adding 10\% or 20\% of  additional fluid results in only a minor change to the $h_c$ and $r_s$ dynamics, independent of whether inward or outward-spreading is considered. Thus, although the ring visibly lifts a meniscus of fluid in the experiments, these simulations suggest that this effect is unlikely to affect spreading dynamics for inward or outward spreading.

\subsubsection{Surfactant annulus}

Because the ring pulls up an annular meniscus of fluid over the span of many minutes, surfactant has time to accumulate at this interface. When the meniscus pinches off, it could therefore leave behind an annular region with a surplus of surfactant. An annulus of surfactant will have the peculiar effect of superimposing both inward and outward spreading at the inner and outer edges of the annulus, respectively. This effect is in addition to the underlying surfactant gradient due to the original inward or outward initial conditions. As will be shown below, this will impact the spreading dynamics.

\begin{figure}
\centering
\includegraphics[height=2in]{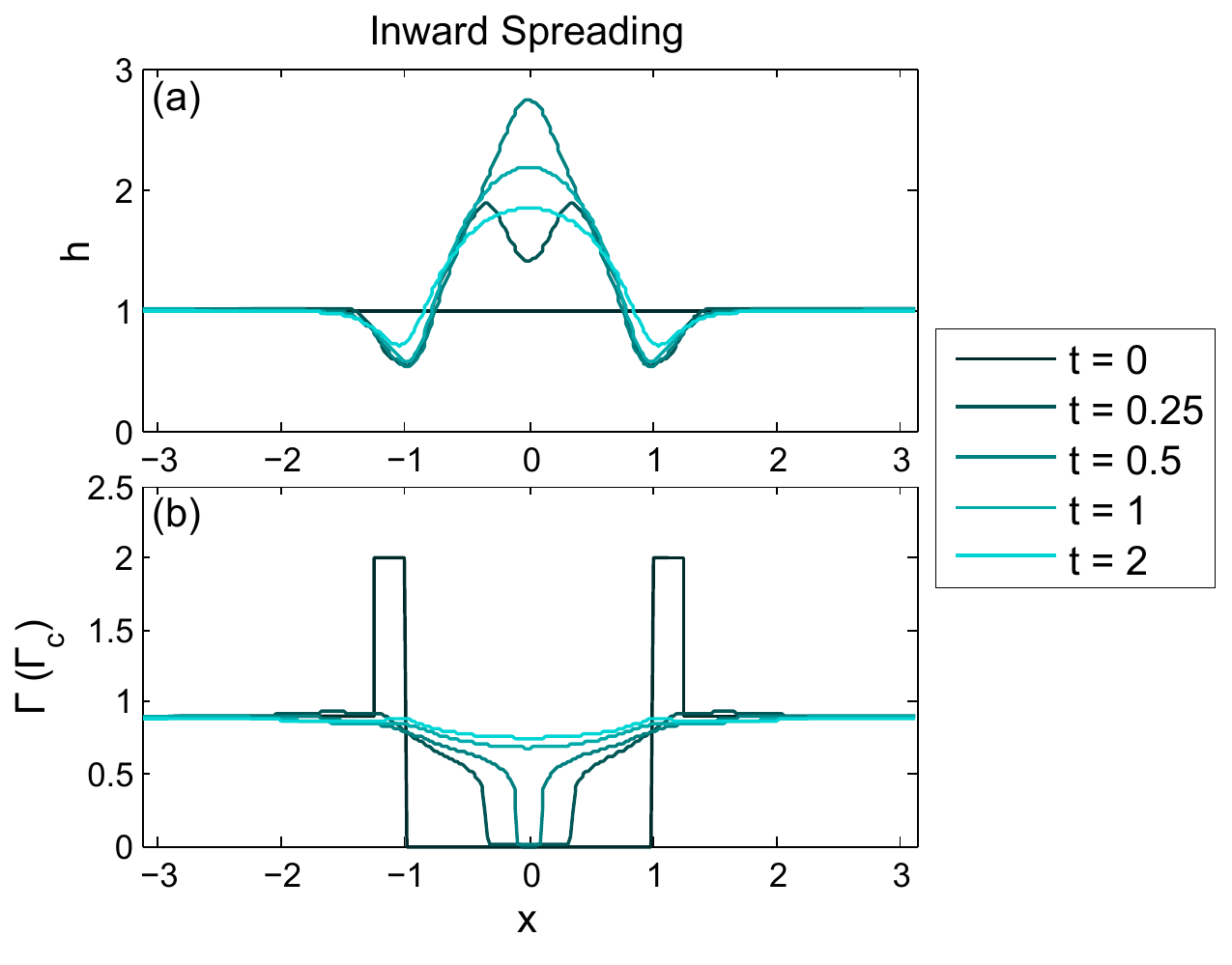}
\includegraphics[height=2in]{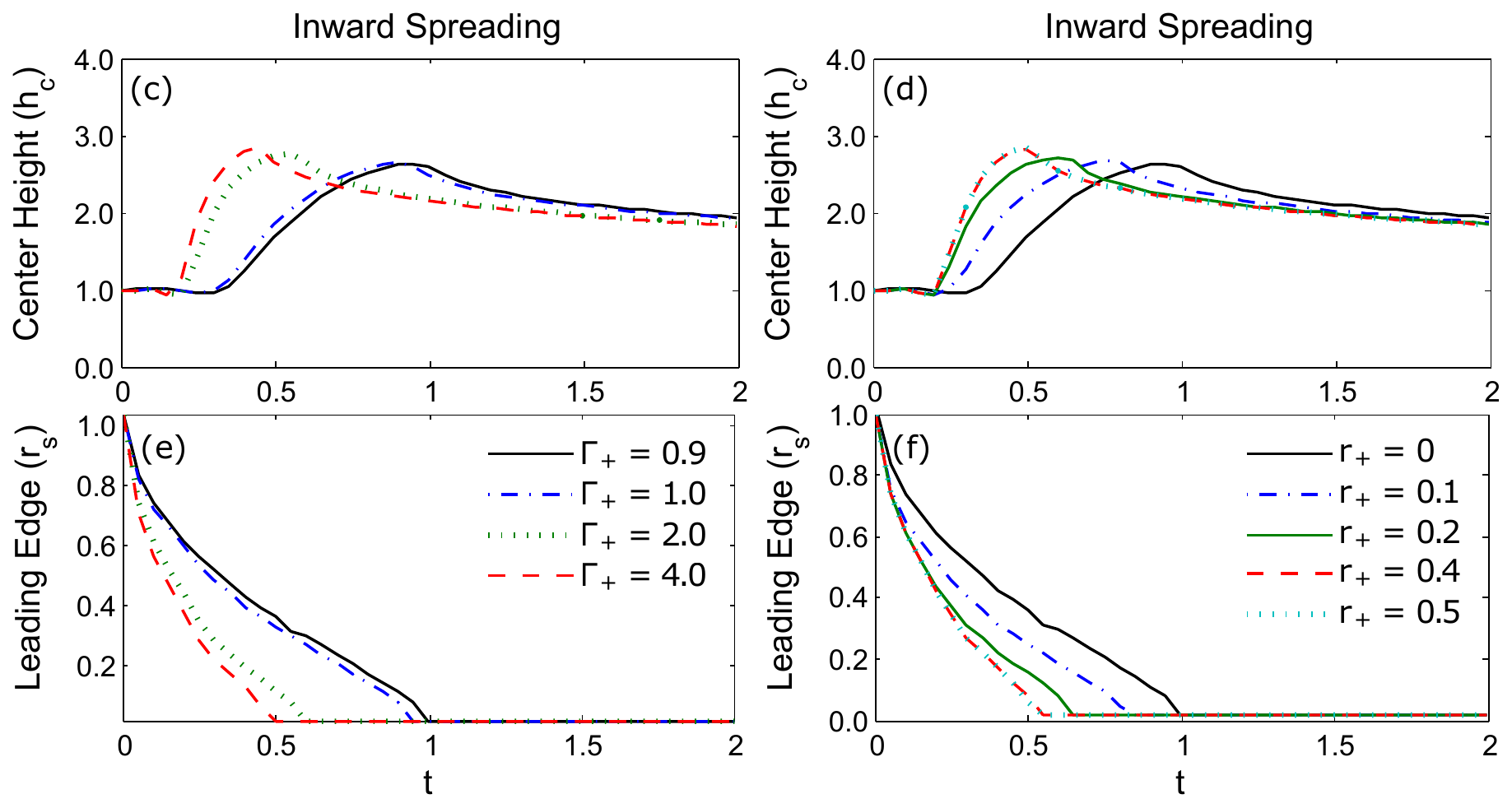}
 \caption{Inward spreading with additional annulus of surfactant at the ring location. 
(a,b) Standard parameters and initial condition (IC4) using a larger initial surfactant concentration ($\Gamma_+ = 2.0$) extending $r_+ = 0.25$ beyond the ring location.
The effect of additional surfactant on $h_c$ and $r_s$ for (c,e) the same parameters, but with varying $\Gamma_+$ and (d,f) the same parameters but varying $r_+$. }
 \label{fig:addpile1}
 \end{figure}

\medskip
\noindent  {\itshape Inward Spreading:} Figure~\ref{fig:addpile1} shows the results of inward spreading simulations obtained by varying both the concentration and width of the annular region.  Increasing the concentration of a fixed annulus width $r_+ = 0.25$ or width of a fixed annulus concentration $\Gamma_+ =2.0$ increases the rate of both growth and decay dynamics.  The larger annulus produces a larger central fluid maximum height $h_c$ at an earlier time in Figure~\ref{fig:addpile1}c and more rapid inward surfactant spreading in Figure~\ref{fig:addpile1}d. With a large annulus concentration, excess surfactant acts as a reservoir (the  EEoS has a very small gradient at large surfactant concentrations).

\begin{figure}
\centering
\includegraphics[height=2.5in]{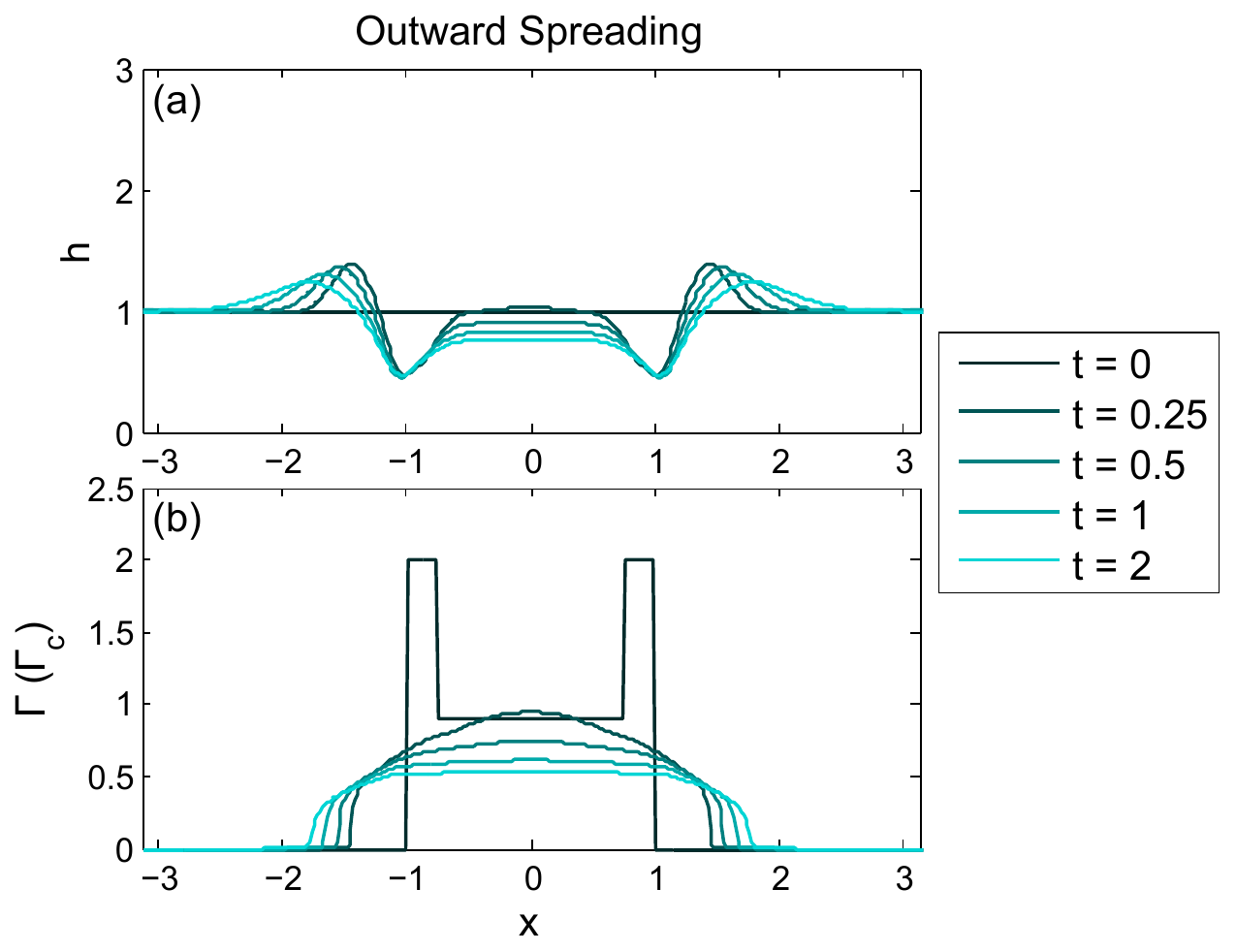}
\includegraphics[height=2.5in]{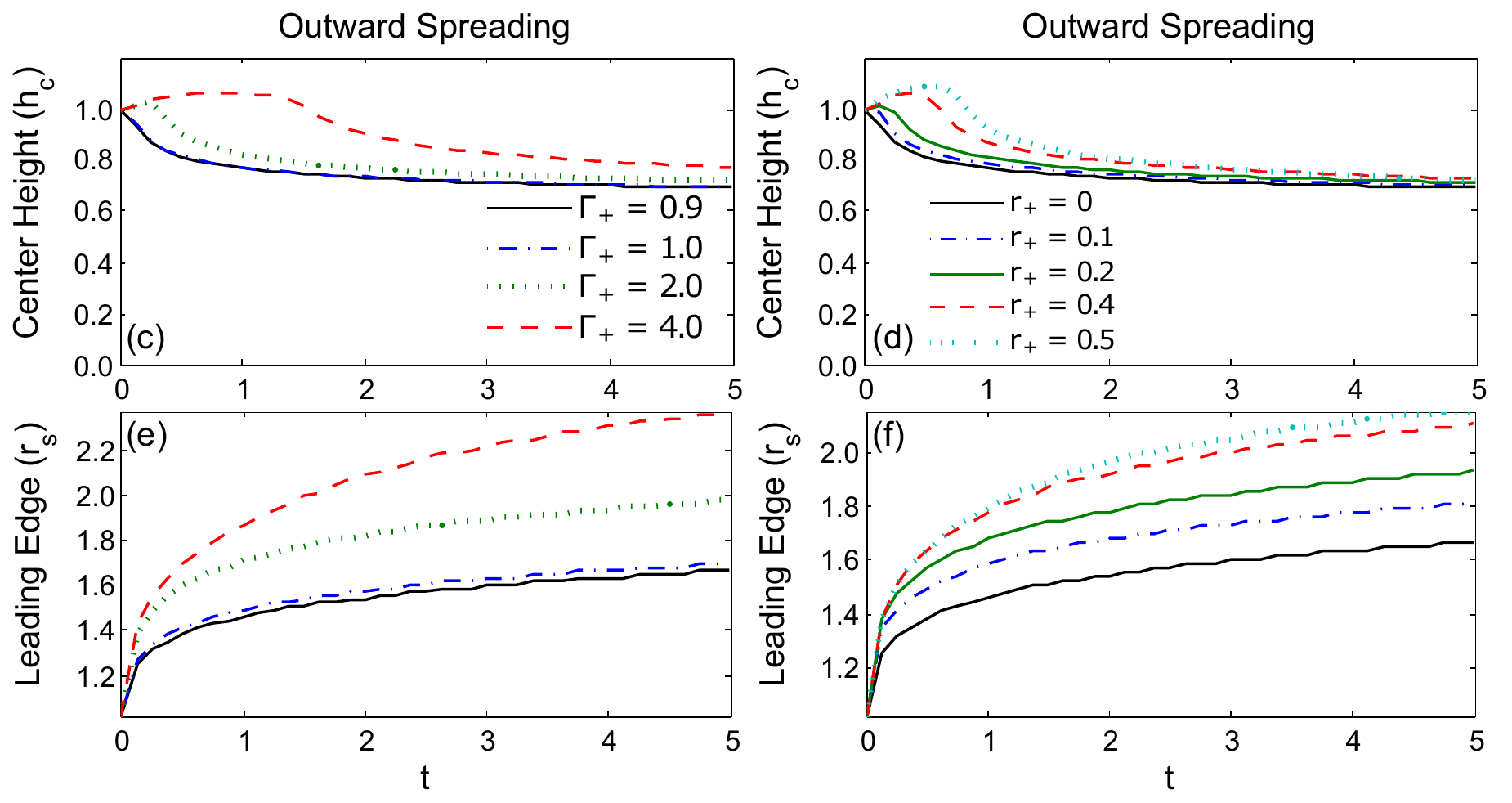}
\caption{The dynamics of the central fluid height $h_c$ and the surfactant leading edge $r_s$ for ouward spreading performed with an additional annulus of surfactant at ring location.  (a,b) Standard parameters and initial condition (IC5).  Plots (c,e) have annulus width $r_+ = 0.25$ and vary $\Gamma_+$.  Plots (d,f) have fixed annulus surfactant concentration $\Gamma_+ = 2.0$  and vary $r_+$.  
}
\label{fig:addpile3}
\end{figure}

\medskip
\noindent {\itshape Outward spreading:}  Figure~\ref{fig:addpile3} illustrates the effect of an annulus of surplus surfactant on the outward spreading dynamics.  As in the inward spreading case, larger surfactant volumes (whether via increased annulus concentration in Figure~\ref{fig:addpile3}a,c or increased width in Figure~\ref{fig:addpile3}b,d) produce a longer relaxation time for the fluid in the center, and more rapid outward spreading of surfactant.  Note that for a large annulus concentration ($\Gamma_+=4.0$), the additional inward spreading is most obvious.  The fluid central fluid height increases as long as the annulus of surfactant is still present, decreasing only once the surplus has spread outward.  Smaller volumes of surfactant equilibrate almost immediately, thus those curves are essentially monotonic while those for larger volumes are not.

%========================================================
\section{Conclusion \label{sec:discuss}}

In this paper, we have tested the ways in which the choice of a realistic equation of state relating surface tension to surfactant concentration influences the outcome of numerical simulations. 
While simplified equations of state have dominated previous studies, including our own,  such as \cite{renardy1996singularly, levy2007gravity,
jensen1992insoluble, jensen1994self, bull2003surfactant, angelini2009bacillus, levy2006motion, jensen1993spreading, warner2004fingering, conti2013effects, braun2012dynamics, witelski2006growing, garcke2006surfactant, barrett2003finite}, 
we find that the spatiotemporal dynamics of surfactant spreading on a thin layer of viscous Newtonian fluid are in fact highly dependent on this choice. Therefore, it is important to incorporate empirical measurements of  $\sigma(\Gamma)$ for the specific materials under investigation. In particular, the correct choice allows for simulations to, for the first time, capture the detailed morphology of a spreading front of surfactant (Investigation 1) which have a distinctive reservoir if the initial surfactant concentration is above $\Gamma_c$.   

We additionally observe that accurately measuring the correct value of $\Gamma_c$ in the equation of state (or picking a different surfactant) will impact predictions for timescales. This effect arises because gradients in $\sigma$ are much stronger for intermediate values of surfactant concentration (above the transition point from the gas phase and below the critical concentration).  The findings from Investigations 1 and 2  indicate that the use of the spreading parameter $S$ is an oversimplification, and other choices should be explored. 

This is particularly important for outward-spreading (as compared with inward-spreading), but only in parameter regimes in which the gradient of $\sigma(\Gamma)$ is strongly affected  (Investigation 2). 

Varying the nondimensional parameters $\kappa$ and $\delta$ does not largely affect the central fluid height evolution nor the surfactant leading edge location for inward or outward spreading.  Varying $\beta$ does impact the dynamics, but the solutions can be scaled to show the similar behavior in the system.    (Investigation 3). 
The lack of sensitivity to $\beta, \kappa, \gamma$ explains why prior attempts to resolve timescale issue by adjusting the non-dimensional parameters have failed. (Even so, as long as the correct values for the fluid are used in the model, there  should be  no flexibility to tune these parameters to better match model and experiment.)  A different nondimensionalization of the timescale could solve some issues with quantitatively predicting spreading rates. 

Finally, the use of empirically-relevant equations of state allows for  simulations to aid experimentalists in addressing how their methods of creating repeatable initial conditions may be impacting the results. We used the empirically-correct equation of state to test for possible artifacts from the use of a retaining ring to generate initial conditions  (Investigation 4). We find that the presence of a fluid meniscus created by such rings will not have an effect on the dynamics. However, if that meniscus also generates a region of surplus surfactant, this can generate long-lived concentration gradients near the original location of the ring, and also affect the velocity of the spreading front. 

These findings will aid in providing improvements in the quantitative agreements between simulations and experiments which have so far been elusive \cite{swanson2014surfactant}. In addition, because they are more time-efficient to perform than laboratory experiments, improved agreement will aid experimentalists by providing a new tool for finding promising new regimes of behavior. 

\section*{Acknowledgments}
This work was funded by  NSF grant DMS-FRG \#096815 (RL and KED), Howard Hughes Medical Institute Undergraduate Science Education Program Award \#52006301 (RL), and Research
Corporation Cottrell Scholar Award \#19788 (RL).  We thank Stephen Strickland for sharing data from experiments, Jonathan Claridge and Jeffrey Wong for collaboration on the code, and Michael Shearer and Ellen Swanson for their contributions regarding the outward spreading system.

%\bibliographystyle{acm}
%\bibliography{InwardSpread,DanielsLab,urls,RayBibExtra}

\begin{thebibliography}{10}

\bibitem{Avanti}
{\sc Avanti Polar Lipids}.
\url{www.avantilipids.com/index.php?option=com_content&view=article&id=1028&Itemid=220&catnumber=810131}

\bibitem{angelini2009bacillus}
{\sc Angelini, T.~E., Roper, M., Kolter, R., Weitz, D.~A., and Brenner, M.~P.}
\newblock Bacillus subtilis spreads by surfing on waves of surfactant.
\newblock {\em Proceedings of the National Academy of Sciences 106}, 43 (2009),
  18109--18113.

\bibitem{barrett2003finite}
{\sc Barrett, J.~W., Garcke, H., and N{\"u}rnberg, R.}
\newblock Finite element approximation of surfactant spreading on a thin film.
\newblock {\em SIAM Journal on Numerical Analysis 41}, 4 (2003), 1427--1464.

\bibitem{borgas1988monolayer}
{\sc Borgas, M.~S., and Grotberg, J.~B.}
\newblock Monolayer flow on a thin film.
\newblock {\em Journal of Fluid Mechanics 193} (1988), 151--170.

\bibitem{braun2012dynamics}
{\sc Braun, R.~J.}
\newblock Dynamics of the tear film.
\newblock {\em Annual Review of Fluid Mechanics 44} (2012), 267--297.

\bibitem{bull2003surfactant}
{\sc Bull, J., and Grotberg, J.}
\newblock Surfactant spreading on thin viscous films: film thickness evolution
  and periodic wall stretch.
\newblock {\em Experiments in Fluids 34}, 1 (2003), 1--15.


\bibitem{ClaridgeGitHubPaper}
{\sc J.~Claridge, R.~Levy, J.~Wong}
\url{https://github.com/claridge/implicit_solvers/blob/master/ImplicitSolversPaper.pdf}

\bibitem{ClaridgeGitHub}
{\sc J.~Claridge, R.~Levy, J.~Wong}
\url{https://github.com/claridge/implicit_solvers/}

\bibitem{conti2013effects}
{\sc Conti, C., Autry, E.~A., Kronmiller, G., and Levy, R.}
\newblock The effects of spatial and temporal grids on simulations of thin
  films with surfactant.
\newblock {\em SIURO 916} (2013), 81--93.

\bibitem{craster2000surfactant}
{\sc Craster, R., and Matar, O.}
\newblock Surfactant transport on mucus films.
\newblock {\em Journal of Fluid Mechanics 425} (2000), 235--258.

\bibitem{craster2009dynamics}
{\sc Craster, R., and Matar, O.}
\newblock Dynamics and stability of thin liquid films.
\newblock {\em Reviews of modern physics 81}, 3 (2009), 1131.

\bibitem{de1994nonlinear}
{\sc De~Wit, A., Gallez, D., and Christov, C.}
\newblock Nonlinear evolution equations for thin liquid films with insoluble
  surfactants.
\newblock {\em Physics of Fluids  6}, 10 (1994), 3256--3266.


\bibitem{glycvisc}
{\sc Dow Chemical}.
\url{http://www.dow.com/optim/optim-advantage/physical-properties/viscosity.htm}

\bibitem{glycdens}
{\sc Dow Chemical}.
\url{http://www.dow.com/optim/optim-advantage/physical-properties/density.htm}


\bibitem{espinosa1993spreading}
{\sc Espinosa, F., Shapiro, A., Fredberg, J., and Kamm, R.}
\newblock Spreading of exogenous surfactant in an airway.
\newblock {\em Journal of Applied Physiology 75}, 5 (1993), 2028--2039.

\bibitem{Fallest-2010-FVS}
{\sc Fallest, D.~W., Lichtenberger, A.~M., Fox, C.~J., and Daniels, K.~E.}
\newblock {Fluorescent visualization of a spreading surfactant}.
\newblock {\em New Journal of Physics 12}, 7 (2010), 73029.

\bibitem{garcke2006surfactant}
{\sc Garcke, H., and Wieland, S.}
\newblock Surfactant spreading on thin viscous films: nonnegative solutions of
  a coupled degenerate system.
\newblock {\em SIAM Journal on Mathematical Analysis 37}, 6 (2006), 2025--2048.

\bibitem{gaver1990}
{\sc Gaver, D.~P., and Grotberg, J.~B.}
\newblock The dynamics of a localized surfactant on a thin film.
\newblock {\em Journal of Fluid Mechanics 213} (1990), 127--148.

\bibitem{Gaver-1992-DST}
{\sc Gaver, D.~P., and Grotberg, J.~B.}
\newblock {Droplet Spreading On A Thin Viscous Film}.
\newblock {\em Journal of Fluid Mechanics 235} (1992), 399--414.

\bibitem{halpern1993surfactant}
{\sc Halpern, D., and Grotberg, J.}
\newblock Surfactant effects on fluid-elastic instabilities of liquid-lined
  flexible tubes: a model of airway closure.
\newblock {\em Journal of Biomechanical Engineering 115}, 3 (1993), 271--277.


\bibitem{jensen1994self}
{\sc Jensen, O.}
\newblock Self-similar, surfactant-driven flows.
\newblock {\em Physics of Fluids  6}, 3 (1994), 1084--1094.

\bibitem{jensen1992insoluble}
{\sc Jensen, O., and Grotberg, J.}
\newblock Insoluble surfactant spreading on a thin viscous film: shock
  evolution and film rupture.
\newblock {\em Journal of Fluid Mechanics 240} (1992), 259--288.

\bibitem{jensen1993spreading}
{\sc Jensen, O., and Grotberg, J.}
\newblock The spreading of heat or soluble surfactant along a thin liquid film.
\newblock {\em Physics of Fluids A: Fluid Dynamics 5}, 1 (1993),
  58--68.

\bibitem{Kaganer1999}
{\sc Kaganer, V., M{\"{o}}hwald, H., and Dutta, P.}
\newblock {Structure and phase transitions in Langmuir monolayers}.
\newblock {\em Reviews of Modern Physics 71}, 3 (1999), 779--819.

\bibitem{karsa1999industrial}
{\sc Karsa, D.}
\newblock {\em Industrial Applications of Surfactants IV}.
\newblock Special publication / Royal Society of Chemistry. Elsevier Science,
  1999.

\bibitem{levy2005partial}
{\sc Levy, R.}
\newblock Partial differential equations of thin liquid films: analysis and  numerical simulation. PhD Thesis, North Carolina State University, 2005.

\bibitem{levy2014pulmonary}
{\sc Levy, R., Hill, D.~B., Forest, M.~G., and Grotberg, J.~B.}
\newblock Pulmonary fluid flow challenges for experimental and mathematical
  modeling.
\newblock {\em Integrative and comparative biology 54}, 6 (2014), 985--1000.

\bibitem{levy2006motion}
{\sc Levy, R., and Shearer, M.}
\newblock The motion of a thin liquid film driven by surfactant and gravity.
\newblock {\em SIAM Journal on Applied Mathematics 66}, 5 (2006), 1588--1609.

\bibitem{levy2007gravity}
{\sc Levy, R., Shearer, M., and Witelski, T.~P.}
\newblock Gravity-driven thin liquid films with insoluble surfactant: smooth
  traveling waves.
\newblock {\em European Journal of Applied Mathematics 18}, 06 (2007),
  679--708.

\bibitem{oron1997long}
{\sc Oron, A., Davis, S.~H., and Bankoff, S.~G.}
\newblock Long-scale evolution of thin liquid films.
\newblock {\em Reviews of Modern Physics 69}, 3 (1997), 931.

\bibitem{otis1993role}
{\sc Otis, D., Johnson, M., Pedley, T., and Kamm, R.}
\newblock Role of pulmonary surfactant in airway closure: a computational
  study.
\newblock {\em Journal of Applied Physiology 75}, 3 (1993), 1323--1333.

\bibitem{pereira2007dynamics}
{\sc Pereira, A., Trevelyan, P., Thiele, U., and Kalliadasis, S.}
\newblock Dynamics of a horizontal thin liquid film in the presence of reactive
  surfactants.
\newblock {\em Physics of Fluids  19}, 11 (2007), 112102.

\bibitem{peterson2010flow}
{\sc Peterson, E.~R.}
\newblock {\em Flow of thin liquid films with surfactant: analysis, numerics,
  and experiment}. PhD Thesis, North Carolina State University,
\newblock 2010.

\bibitem{peterson2011radial}
{\sc Peterson, E.~R., and Shearer, M.}
\newblock Radial spreading of a surfactant on a thin liquid film.
\newblock {\em Applied Mathematics Research eXpress 2011}, 1 (2011), 1--22. 

\bibitem{Reis2009}
{\sc Reis, P., Holmberg, K., Watzke, H., Leser, M.~E., and Miller, R.}
\newblock {Lipases at interfaces: A review}.
\newblock {\em Advances in Colloid and Interface Science 147–148\/} (2009),
  237--250.

\bibitem{renardy1996singularly}
{\sc Renardy, M.}
\newblock A singularly perturbed problem related to surfactant spreading on
  thin films.
\newblock {\em Nonlinear Analysis: Theory, Methods \& Applications 27}, 3
  (1996), 287--296.

\bibitem{strickland2015surfactant}
{\sc Strickland, S.~L.}
\newblock {Surfactant dynamics: Spreading and wave induced dynamics of a monolayer. PhD Thesis, North Carolina State University (2015).}

\bibitem{Strickland2014}
{\sc Strickland, S.~L., Hin, M., Sayanagi, M.~R., Gaebler, C., Daniels, K.~E.,
  Levy, R., and Conti, C.}
\newblock {Self-healing dynamics of surfactant coatings on thin viscous films}.
\newblock {\em Physics of Fluids 26}, 4 (Apr. 2014), 042109.

\bibitem{swanson2014surfactant}
{\sc Swanson, E.~R., Strickland, S.~L., Shearer, M., and Daniels, K.~E.}
\newblock Surfactant spreading on a thin liquid film: reconciling models and
  experiments.
\newblock {\em Journal of Engineering Mathematics\/} (2014), 1--17.

\bibitem{tiberg1994spreading}
{\sc Tiberg, F., and Cazabat, A.-M.}
\newblock Spreading of thin films of ordered nonionic surfactants. origin of
  the stepped shape of the spreading precursor.
\newblock {\em Langmuir 10}, 7 (1994), 2301--2306.

\bibitem{warner2004fingering}
{\sc Warner, M., Craster, R., and Matar, O.}
\newblock Fingering phenomena associated with insoluble surfactant spreading on
  thin liquid films.
\newblock {\em Journal of Fluid Mechanics 510\/} (2004), 169--200.

\bibitem{witelski2006growing}
{\sc Witelski, T.~P., Shearer, M., and Levy, R.}
\newblock Growing surfactant waves in thin liquid films driven by gravity.
\newblock {\em Applied Mathematics Research Express 2006\/} (2006), 15487.

\end{thebibliography}

\end{document}